\documentclass[fleqn,usenatbib,article]{mnras}

\usepackage{newtxtext,newtxmath}

\usepackage[T1]{fontenc}

\DeclareRobustCommand{\VAN}[3]{#2}
\let\VANthebibliography\thebibliography
\def\thebibliography{\DeclareRobustCommand{\VAN}[3]{##3}\VANthebibliography}

\usepackage{graphicx}	
\usepackage{amsmath}	
\usepackage{amssymb}	
 
\usepackage{multirow}
\usepackage{subfigure}

\title[FRB energy and redshift distributions]{On the energy and redshift distributions of fast radio bursts}

\author[Zhang et al.]{
Rachel C. Zhang$^{1}$\thanks{E-mail: rczhang@mit.edu},
Bing Zhang$^{2}$,
Ye Li$^{3,4}$
and Duncan R. Lorimer$^{5,6}$
\\
$^{1}$Department of Physics, Massachusetts Institute of Technology, 77 Massachusetts Avenue, Cambridge, MA, 02139, USA\\
$^{2}$Department of Physics and Astronomy, University of Nevada Las Vegas, 4505 S. Maryland Parkway, Las Vegas, NV, 89154, USA\\
$^{3}$Kavli Institute for Astronomy and Astrophysics, Peking University, Beijing 100871, China \\
$^{4}$Purple Mountain Observatory, Chinese Academy of Sciences, Nanjing 210008, China \\
$^{5}$Department of Physics and Astronomy, West Virginia University, P.O. Box 6315, Morgantown, WV, USA \\
$^{6}$ Center for Gravitational Waves and Cosmology, West Virginia University, Chestnut Ridge Research Building, Morgantown, WV, USA
}

\date{Accepted XXX. Received YYY; in original form ZZZ}

\pubyear{2015}

\begin{document}
\label{firstpage}
\pagerange{\pageref{firstpage}--\pageref{lastpage}}
\maketitle

\begin{abstract}
Fast radio bursts (FRBs) are millisecond-duration radio transients from cosmological distances. Their isotropic energies follow a power-law distribution with a possible exponential cutoff, but their intrinsic redshift distribution, which contains information about the FRB sources, is not well understood. We attempt to constrain both distributions by means of Monte Carlo simulations and comparing the simulations results with the available FRB specific fluence distribution, dispersion measure (DM) distribution, and the estimated energy distribution data. Two redshift distribution models, one tracking the star formation history of the universe and another tracking compact binary mergers, are tested. For the latter model, we consider three merger delay timescale distribution (Gaussian, log-normal, and power law) models. Two FRB samples detected by Parkes and the Australian Square Kilometre Array Pathfinder (ASKAP), respectively, are used to confront the simulation results. We confirm the $\sim -1.8$ power law index for the energy distribution but the exponential cutoff energy of the distribution, if any, is unconstrained. For the best energy distribution model, none of the redshift distributions we considered are rejected by the data. A future, larger, uniform FRB sample (such as the one collected by the Canadian Hydrogen Intensity Mapping Experiment, CHIME) can provide better constraints on the intrinsic FRB redshift distribution using the methodology presented in this paper.
\end{abstract}




\section{Introduction}

Fast radio bursts (FRBs) are millisecond-duration transients in the radio band \citep{lorimer07,thornton13,petroff19,cordes19}. In recent years, they have been detected by many radio telescopes across the globe, e.g.~Parkes, the Australian Square Kilometre Array Pathfinder (ASKAP), the Canadian Hydrogen Intensity Mapping Experiment (CHIME), Green Bank Telescope (GBT), the Five-hundred-meter Aperture Spherical radio Telescope (FAST), and the second Survey for Transient Astronomical Radio Emission (STARE2). The rapid growth of the detected FRB events allows one to study the physical mechanism of FRB sources using the statistical properties of FRBs.

Two key intrinsic distributions of FRB sources shape the observed FRB population. The first is the isotropic-equivalent energy (or luminosity) of the bursts. The second is the intrinsic redshift distribution of FRB sources, which is closely related to the largely unknown astrophysical sources that produce cosmological FRBs.

Concerning the energy distribution, independent studies by different groups \citep[e.g.][]{luo18,luo20,lu19b,lu20b} suggested that the FRB energy (luminosity) distribution is consistent with a power law with an index $\sim -1.8$. This power law covers at least eight orders of magnitude ranging from $\sim 10^{35}$---$10^{43}$~erg ($\sim 10^{38}$---$10^{46}~{\rm erg \ s^{-1}}$) \citep{lu20}. With a sample of 46 FRBs, \cite{luo20} suggested that there is an exponential cutoff  around $3\times 10^{44} \ {\rm erg \ s^{-1}}$ in the luminosity function, which corresponds to a cutoff of $\sim 3\times 10^{41}$~erg in the energy distribution. 

\begin{figure*}
	\includegraphics[width=0.8\paperwidth]{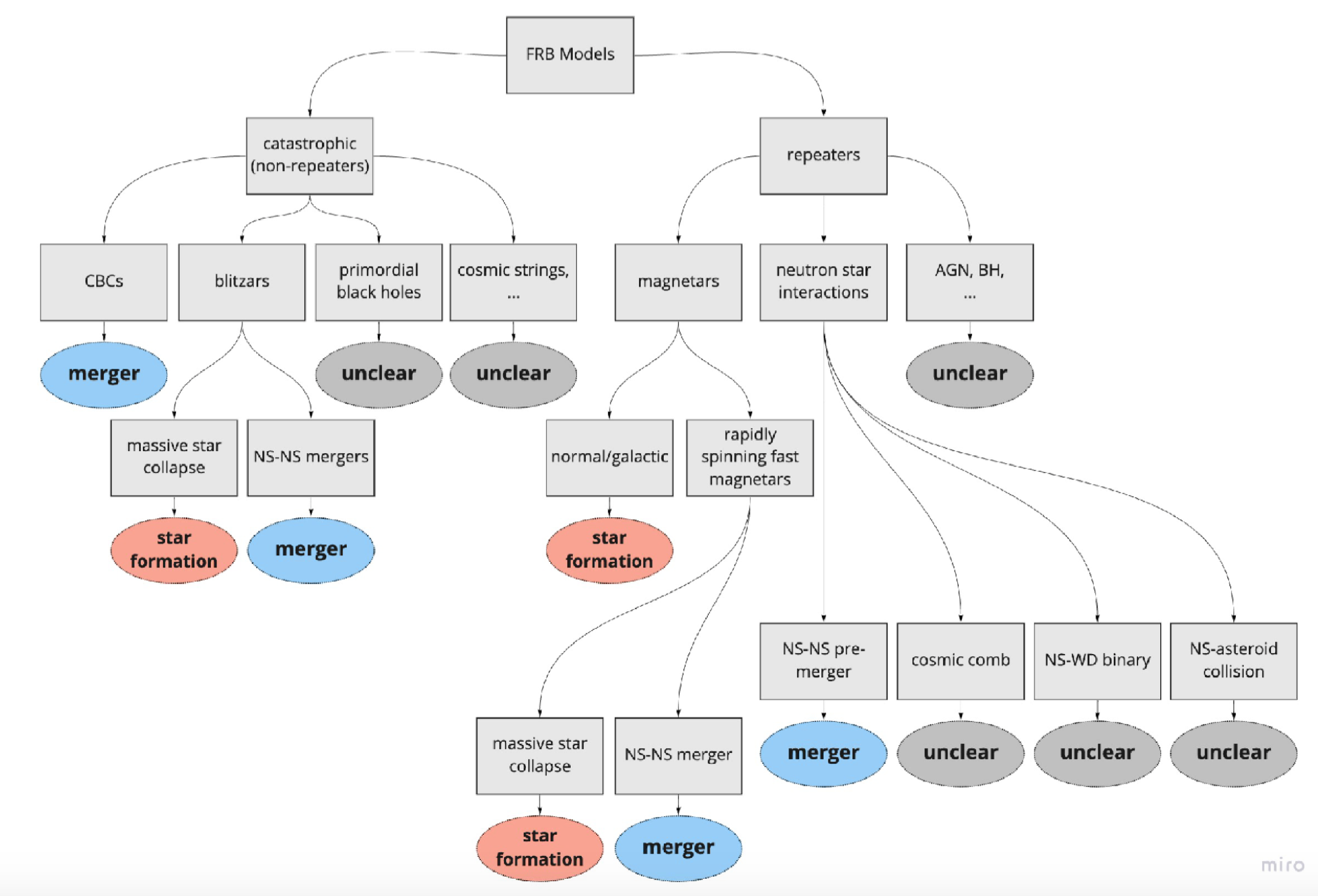}
    \caption{A flow chart for the various redshift distributions implied by different FRB models.}
    \label{fig:figure1}
\end{figure*}

 As shown in Fig.~\ref{fig:figure1}, different FRB source models predict different redshift distributions. In general, FRB models can be divided into catastrophic (non-repeater) and repeater models. Since the FRB event rate density exceeds that of catastrophic events by a large margin \citep{ravi19b,luo20}, the majority of FRBs likely originate from repeating sources. The leading model for the repeaters involve magnetars \citep{popov10,katz16,murase16, metzger17,beloborodov17,kumar17,yangzhang18,metzger19,beloborodov20,wadiasingh20}, neutron stars with super strong magnetic fields. This model becomes especially popular after the recent discovery of FRB~20200428 in association with a hard X-ray burst from a Galactic magnetar \citep{STARE2-SGR, CHIME-SGR, HXMT-SGR, Integral-SGR, Konus-SGR, AGILE-SGR}. Since most magnetars are believed to be produced from recent supernova explosions, this model predicts that the FRB rate follows the cosmic star formation rate. One special channel to form rapidly spinning magnetars invokes mergers of binary neutron stars (BNSs) \citep[e.g.][]{margalit19,wang20}\footnote{Such a scenario requires a  stiff neutron star equation of state and a large maximum neutron star mass $M_{\rm TOV}$, which is inconsistent with the suggestion that the merger product of GW~170817 is a black hole \citep[e.g.,][]{margalit17}, but is consistent with interpreting X-ray plateaus following most short GRBs as emission from a supramassive or stable magnetar \citep[e.g.][]{gao16}.}. Since it takes an extra delay timescale for a neutron star binary to merge due to gravitational wave radiation, this channel predicts a very different redshift distribution from the SFR one, which has been widely discussed within the context of short-duration gamma-ray bursts (GRBs) \citep[e.g.][]{virgili11,wanderman15,sun15}. 

Another repeating FRB model that invokes such a redshift distribution is the repeated magnetosphere interaction model for BNSs decades to centuries before the mergers \citep{zhang20}. If such a channel dominates the FRB population, then the majority of the FRBs should follow a redshift distribution that track compact star mergers. Besides, there are other suggested repeater models (e.g., models invoking AGNs \citep{romero16}, white dwarfs \citep{gu16}, interactions between neutron stars and ram pressure of a nearby astrophysical stream -- the so-called ``cosmic comb'' model \citep{zhang17}, or interactions between asteroids/comets and neutron stars \citep{dai16}). The redshift distributions of these models are not well defined, but is likely between the two extreme models discussed above: the SFR model and the merger model that invokes a significant delay from the SFR history. 

Finally, if there is indeed a significant population of catastrophic FRBs, they will introduce additional redshift distribution mixes. The compact binary coalescence (CBC) induced FRBs \citep[e.g.][]{totani13,wang16,zhang16a,levin18} would likely follow the BNS merger redshift distribution. The so-called ``blitzars'' \citep{falcke14} may either track the SFR history (for supramassive neutron stars born from massive star core collapse) or BNS merger history (for BNS-merger generated supramassive neutron stars) \citep{zhang14}. Other more exotic channels, e.g. primordial black hole evaporation \citep{keane12} or cosmic strings \citep{vachaspati08,yu14}, would introduce a different redshift distribution, which is not well quantified. Notice that these catastrophic models can at most account for a small fraction of observed FRBs, so that if these progenitors indeed exist they can only account for a sub-category of FRBs. Redshift distribution constraints alone cannot help to identify or rule out the existence of these progenitors.

A handful of FRBs have been localized, leading to direct measurements of their redshifts from their host galaxies \citep{tendulkar17,bannister19,ravi19,prochaska19,macquart20}. This small sample is not adequate to give a constraint on the redshift distribution of FRBs. On the other hand, it confirms the long expected relationship between redshift $z$ and the dispersion measurement (DM) after deducting the contribution from the Milky Way galaxy \citep{macquart20,lizx20}. This raises the possibility of constraining the redshift distribution of FRBs using the DM data which are available for a much larger sample of FRBs. The host galaxy types and the offsets of the FRB locations from the host galaxies are helpful to infer the possible FRB sources. However, current data are consistent with both core collapse supernovae and compact star mergers \citep{lizhang20,heintz20,bochenek20b} and therefore cannot constrain FRB redshift distribution.

So far, the redshift distribution of FRBs is not well constrained. It is commonly assumed that it follows the SFR history, but it has been claimed that a test on 22 FRB events detected by Parkes favors a BNS merger redshift distribution model invoking a power law distribution of the delay timescale \citep{cao18}. \cite{locatelli19} performed the $\left< V/V_{\rm max} \right>$ test for FRBs and claimed that FRBs may have a redshift evolution faster than star formation history. However, no detailed Monte Carlo simulations have been carried out to confront a larger data sample available from the FRB catalogue \citep{petroff16} with an array of redshift distribution models. This is the subject of this paper. We present the models in Section \ref{sec:models}. Observational data and Monte Carlo simulations are described in Section \ref{sec:data&simulations}.The results are presented in Section \ref{sec:results}. Our main conclusions and further discussion are presented in Section \ref{sec:conclusions}.

\section{Models}\label{sec:models}

\subsection{Energy Distribution}
\label{energy}
The luminosity function of FRBs, which is the event rate per unit cosmic co-moving volume per unit luminosity, has been constrained by several groups as a power law  \citep{luo18, luo20, lu19, lu20b, lu20, fialkov18}, probably with an exponential cutoff at the high end \citep{luo20}\footnote{Earlier speculation of a log-normal luminosity function \citep[e.g.][]{caleb16} is no longer supported by the data {\citep{luo18, luo20, fialkov18, lu19b, lu20b}}.}. Since the observed durations of FRBs are typically of the order of milliseconds, this luminosity function may be translated to an isotropic energy distribution in the form of

    \begin{equation}
        \frac{dN}{dE} \propto \left(\frac{E}{E_c}\right)^{-\alpha} e^{-\frac{E}{E_c}}, 
    \label{eq1}
    \end{equation}
where $E_c$ is the exponential cutoff energy. 
Previous authors showed that $\alpha \sim 1.8$ \citep{luo18,luo20,lu19,lu20b} is consistent with the data. \cite{luo20} suggested a luminosity cutoff at $L_c \sim 3\times 10^{44} \ {\rm erg \ s^{-1}}$, which corresponds to $E_c \sim 3 \times 10^{41}$ erg.

\subsection{Redshift Distributions}\label{sec:z-dis}

In general, the redshift distribution of FRBs can be described by the intrinsic event rate density distribution $\frac{dN}{dtdV}$ as a function of redshift, $z$. Once this function is known, the observed FRB number per unit time per unit redshift bin can be derived by
        \begin{equation}
            \frac{dN}{dt_{\rm obs}dz} = \frac{dN}{dtdV} \cdot \frac{dt}{dt_{\rm obs}} \cdot \frac{dV}{dz},
            \label{eq:dN/dtdz}
        \end{equation}
        where 
        \begin{equation}
            \frac{dt}{dt_{\rm obs}} = \frac{1}{1+z}
        \end{equation}
        due to the cosmological time dilation effect, and the redshift-dependent specific comoving volume is
        \begin{equation}
            \frac{dV}{dz} = \frac{c}{H_0} \cdot \frac{4 \pi D_{\rm L}^2}{(1+z)^2 \sqrt{\Omega_m(1+z)^3+\Omega_\Lambda}}.
        \end{equation}
        Here, $D_{\rm L}$ is the luminosity distance, $c$ is speed of light, $H_0$ is Hubble constant, and $\Omega_m$ and $\Omega_\Lambda$ are density fractions for matter and dark energy, respectively.

\subsubsection{Star Formation Rate History Model}
\label{sfh}
The SFR history of the universe has been studied by many authors \citep{hopkins06, madau14, yuksel08}. We adopt a model by \cite{yuksel08} who fit a wide range of SFR data using an empirical formula
        \begin{equation}
            \label{eq:SFR}
            \frac{dN}{dtdV} \propto \left[(1+z)^{a\eta}+\left(\frac{1+z}{B}\right)^{b\eta}+\left(\frac{1+z}{C}\right)^{c\eta}\right]^{1/\eta}
        \end{equation}
        where $a = 3.4, b = -0.3, c = -3.5, B \simeq 5000, C \simeq 9,$ and $\eta = -10$.
This empirical model has three redshift segments and arguably can catch more detailed features in the star formation history. If FRBs follow SFR history, we apply Eq.~(\ref{eq:SFR}) to predict its redshift distribution.

\subsubsection{Compact Star Merger Model}
\label{sec:merger}
In merger events, a binary system needs to undergo a long inspiral phase before the final coalescence, so there is an additional time delay with respect to the creation of the system. The distribution of this time delay timescale also plays an important role in defining the final $\frac{dN}{dtdV}$ events. In the literature, three types of merger delay timescale $\tau$ distributions have been discussed \citep{virgili11,wanderman15,sun15}:
\begin{itemize}
    \item For a Gaussian delay timescale model, the probability distribution function for $\tau$ is
    \begin{equation}
        P(\tau) d\tau = \frac{1}{\sqrt{2\pi} \sigma} \exp \left( - \frac{(\tau-\tau_0)^2}{2 \sigma^2} \right)  d\tau,
    \end{equation}
    where $\tau_0 = 2$ Gyr, and $\sigma = 0.3$ Gyr for the best-fitting parameters \citep{virgili11};
    \item For a log-normal delay timescale model, the probability distribution function for $\tau$ is
    \begin{equation}
        P(\tau) d\ln\tau = \frac{1}{\sqrt{2\pi} \sigma} \exp \left( - \frac{(\ln\tau-\ln\tau_0)^2}{2 \sigma^2} \right)  d\ln\tau,
    \end{equation}
    where $\tau_0 = 2.9$ Gyr, and $\sigma = 0.2$ for the best-fitting parameters \citep{wanderman15};
    \item For a power-law delay timescale model, the probability distribution function for $\tau$ is
    \begin{equation}
        P(\tau) d\tau = \left(\frac{1-\alpha}{\tau_{\rm max}^{1-\alpha_{\tau}}-\tau_{\rm min}^{1-\alpha_{\tau}}}\right)
        \tau^{-\alpha_\tau} d \tau,
    \end{equation}
    where  $\tau_{\rm max}$ and $\tau_{\rm min}$ represent the maximum and minimum merger delay timescale, respectively, and $\alpha_\tau = 0.81$ for the best-fitting parameter \citep{wanderman15}. 
\end{itemize}
Testing these models against short GRB data suggest that the power law model is disfavoured while the other two models are consistent with the data \citep{virgili11,wanderman15,sun15}. For general purposes, we consider all three models in this paper. 

For the compact star merger model for FRBs, to obtain the intrinsic FRB event rate density distribution $\frac{dN}{dtdV}$ as a function of redshift $z$, we take the following steps through Monte Carlo simulations: 
    \begin{enumerate}
        \item Convert the SFR history redshift distribution into a star formation history lookback time distribution. For a redshift $z$, the lookback time is defined as 
        \begin{equation}
            t_{\rm L} = \int_0^z{\frac{t_{\rm H}}{(1+z)\sqrt{\Omega_m(1+z)^3+\Omega_\Lambda}}}dz,
        \end{equation}
        where $t_{\rm H}=1/H_0$ is the Hubble time. 
        \item Adopt one of the three merger delay timescale models and generate a delay timescale distribution.
        \item For each FRB, subtract the merger delay times from the star formation history lookback time, and obtain the new lookback timescale for the merger event. If it is greater than zero, keep it in the simulated sample. Otherwise (which means that the merger happens in the future), drop out the simulation point from the sample.
        \item Convert the final lookback time back into the redshift, and derive the new event rate density distribution $\frac{dN}{dtdV}$ for the merger sample. 
        \item Convert 
        $\frac{dN}{dtdV}$ into $ \frac{dN}{dt_{\rm obs}dz}$ using Eq.~(\ref{eq:dN/dtdz}).
    \end{enumerate}

\subsection{Specific Fluence and Observed Sample} 

For a mock FRB with isotropic energy $E$ and redshift $z$, assuming a flat radio spectrum, one can calculate its specific fluence as detected from Earth \citep{zhang18a}
    \begin{equation}
        {\cal F}_\nu = \frac{(1+z)E}{4\pi D_{\rm L}^2 \nu_c},
    \end{equation}
where $\nu_c$ is the central observing frequency. To obtain the mock ``observed'' sample, we need to introduce a specific fluence sensitivity threshold ${\cal F}_{\rm \nu,th}$, which is taken as a free parameter to allow us to mimic detections by different radio telescopes.

\subsection{Dispersion Measure} 
To confront mock FRBs with observed FRBs, one needs to connect $z$ with the measured quantity dispersion measure (DM). In general, the observed DM has several components, i.e. $\rm DM=DM_{MW}+DM_{IGM}+DM_{host}/ (1+{\it z})$, where $\rm DM_{MW}$, $\rm DM_{IGM}$, and ${\rm DM_{host}}$ are the contributions from the Milky Way, intergalactic medium (IGM), and FRB host, respectively \citep{ioka03,inoue04,deng14}. For fully ionized hydrogen and helium (which is true for the redshift range from which FRBs are detected), the second component $\rm DM_{IGM}$ is directly connected to $z$ through \citep{deng14,macquart20,lizx20}
    \begin{equation}
        {\rm DM}_{\rm IGM} (z) = \frac{3cH_0\Omega_bf_{\rm IGM}}{8\pi Gm_p}\int_0^z{\chi\frac{1+z}{\sqrt{\Omega_m(1+z)^3+\Omega_\Lambda}}} dz,
        \label{eq:DMIGM}
    \end{equation}
    where $\chi = {7}/{8}$, 
    $m_p = 1.673 \times 10^{-27}$ g is the proton mass, and $f_{\rm IGM} = 0.84$ is the fraction of baryons in the IGM.

Observationally, $\rm DM$ and $\rm DM_{MW}$ are directly measured or obtained from Milky Way electron density models \citep{cordes02,yao17}. It is therefore convenient to define the extragalactic DM
 \begin{equation}
        {\rm DM_E} = {\rm DM_{\rm IGM}}(z) + \frac{{\rm DM_{\rm host}}}{1+z},
        \label{eq:DME}
    \end{equation}
which can be directly compared with the model results. 
    
\section{Data and Simulations}\label{sec:data&simulations}

\subsection{Data}
We use the publicly available FRB catalog from \cite{petroff16}\footnote{Available at the website \url{http://www.frbcat.org}.}. As of October 2020, there are a total of 118 FRBs in the catalog. However, in order to keep consistency in threshold cutoffs from simulations, we need to select sub-samples of FRBs detected by the same telescopes. We choose two subsamples from the catalog with the greatest number of FRB events detected by the same telescopes, i.e., the ASKAP \citep[e.g.][]{bannister17,shannon18,macquart19} and Parkes \citep[e.g.][]{petroff16} samples. Both samples have 27 events each. We extract from the events the Milky Way DM, DM, width, specific flux, and center frequency, to calculate the excess DM, $\rm DM_E$ = DM - $\rm DM_{MW}$, specific fluence: specific flux $\cdot$ width, and hence energy: specific fluence $\cdot$ $4\pi D_{\rm L}^2$ $\cdot$ center frequency divided by $(1+z)$. These derived quantities are compared with the simulation results as described below.

\subsection{Simulations}
\begin{figure}
	\includegraphics[width=\columnwidth]{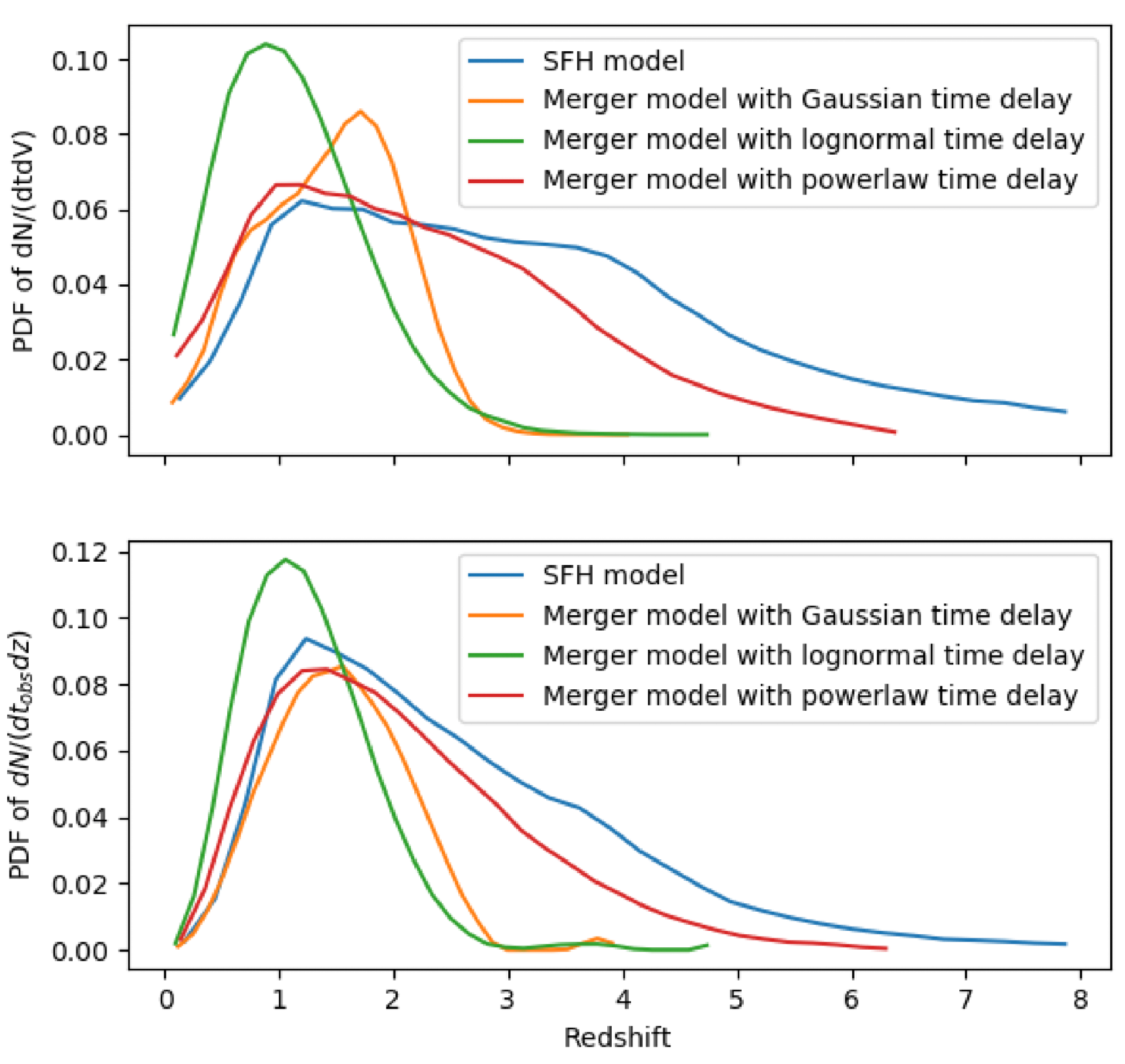}
    \caption{Probability distribution functions for the star formation rate history (SFR) model and three merger models with different merger delay timescale distribution models. The top panel shows the intrinsic FRB event rate density distribution $dN/(dtdV)$, while the bottom panel shows the observed FRB event rate redshift distribution $dN/(dt_{\rm obs}dz)$. }
    \label{fig:fig2}
\end{figure}

We conduct Monte Carlo simulations to generate mock FRBs with redshifts in the range 0--8 drawn from the distributions
as described in Section \ref{sec:z-dis}. For each of the four $z$-distribution models (SFR model, Gaussian merger, lognormal merger, and power law merger models), we generate 200,000 mock FRBs. We choose 200,000 to make sure that we have a large enough number of FRBs after the fluence threshold cut to perform statistics with sufficient significance. The parameters of each model are adopted based on the best-fitting parameters obtained from short GRB studies \citep{virgili11,wanderman15,sun15}. We follow the steps outlined in \ref{sec:merger} to get the proper $\frac{dN}{dtdV}$ distribution. For all four models, we use Eq.(\ref{eq:dN/dtdz}) to obtain their respective intrinsic redshift distributions. Fig.~\ref{fig:fig2} presents the $\frac{dN}{dtdV}$ distributions (upper panel) and the $\frac{dN}{dt_{\rm obs}dz}$ distributions (lower panel) of all four models from our simulations. One can see that the SFR model has the widest $z$-distribution. Because of the merger delay, all the merger models tend to have narrower $z$ distributions with the overall redshift progressively shifted towards lower redshifts following the order of power law, Gaussian, and lognormal models. 

Each mock FRB is also assigned an associated energy randomly chosen from a distribution following Eq.(\ref{eq1}). Combined with the corresponding redshift distribution, one can obtain a specific fluence distribution of the mock sample. To obtain a mock ``observed'' sample, we manually set a specific fluence cutoff for a particular telescope. For the ASKAP and Parkes samples, we adopt $26 \ \rm Jy \ ms$ \citep{james19b} and $2 \ \rm Jy \ ms$ \citep{keane15} as the sensitivity threshold, respectively. 

Depending on the threshold specific fluence, we choose different energy ranges to perform simulations. This is because a higher threshold telescope (e.g., ASKAP) would not be able to detect low-energy events at cosmological distances, and therefore it makes no sense to simulate very low-energy FRBs. We try different lower bound energies in the simulations from $10^{37}$ erg in increments of half order-of-magnitude and find the optimum value of the lower energy to most efficiently obtain enough ``observed'' mock samples without missing low-energy events. For the Parkes telescope sample, we choose the energy range from $10^{37.5}$ to $10^{44}$~erg, while for the ASKAP telescope sample, we change the lower bound to $10^{38.5}$~erg. For the power law index, since $\alpha$ has been constrained to be around 1.8 with uncertainties, we test three values: 1.6, 1.8, 2.0, all with a cutoff energy $E_c = 10^{41.5}$ erg. For the case of $\alpha=1.8$, we test three cutoff energies:
$E_c = 10^{41.5}, 10^{42.5}, 10^{43.5}$ erg, respectively. 

Throughout the paper, we care about the energy distribution rather than luminosity distribution so we do not simulate the durations (widths) of FRBs. This is because the FRB detection sensitivity threshold is often denoted in terms of specific fluence (Jy ms) rather than specific flux. Since the observed FRB width distribution clusters around a millisecond \citep{chime-repeaters}, the constrained energy distribution can be roughly translated to a constraint on the luminosity function by dividing the energy by a characteristic width $\sim 1$ ms. We also do not simulate the FRB spectrum by assuming that the energy in investigation is the energy deposited around GHz band. This simplification is justified when applied to the ASKAP and Parkes telescopes, both with the central frequencies around GHz: 1.297 or 1.320~GHz for ASKAP and 1.352~GHz for Parkes.

With simulated mock FRB samples, we calculate the simulated $\rm DM_E$ distribution using Eq.~(\ref{eq:DME}). The IGM contribution, $\rm DM_{IGM}$ is calculated directly from Eq.~(\ref{eq:DMIGM}). It is known that the true $\rm DM_{IGM}$ for a particular $z$ could deviate from the value of Eq.~(\ref{eq:DMIGM}) due to density fluctuation of large-scale structure \citep[e.g.][]{mcquinn14}. However, since we are simulating a large sample of mock bursts and only care about the distributions of various parameters, using the central value given by Eq.~(\ref{eq:DME}) is adequate. Similarly, we also use the central value of $\rm DM_{host}$, $107 \ {\rm pc \ cm^{-3}}$ as constrained from data  \citep{lizx20} to calculate $\rm DM_E$ of the mock samples in Eq.~(\ref{eq:DME})\footnote{The number $107 \ {\rm pc \ cm^{-3}}$ is derived from fitting the available data by \cite{lizx20}. Because the data sample is small, it is difficult to constrain the redshift evolution of $\rm DM_{host}$ directly from the data. Theoretically, one would expect that the average $\rm DM_{host}$ becomes smaller at larger redshifts. However, since the DM contribution from the host is smaller by a factor $(1+z)$, this contribution is in any case small, especially at high redshifts. Also in view of other uncertainties in the DM estimates, for simplicity, we used a simplified model without introducing the possible $z$-evolution in $\rm DM_{host}$.}.

\section{Results}\label{sec:results}
\subsection{One-Dimensional KS Test on Specific Fluence, $\rm \bf DM_E$, and Energy}
    \subsubsection{Method}
    The 1-D Kolmogorov-Smirnov (KS) test compares the cumulative distribution functions (CDFs) $P(<x)$ and $P(<x')$ of two distributions $x$ and $x'$. The two distributions are often a theoretical model and a data sample or two data samples. This test is done to determine the probability of both distributions coming from the same inherent distribution by finding the largest distance between the two CDFs $D_{\rm KS}$ = max($P(<x) - P(<x')$). 
    \subsubsection{Results}
    We apply the 1-D KS test to compare the specific fluence, $\rm DM_E$, and energy distributions between an FRB data sample and a simulated ``observed'' FRB sample. We confront the two data samples (Parkes and ASKAP) each with four simulated model samples: star formation rate history (SFR), merger with Gaussian delay ($\rm M_G$), merger with lognormal delay ($\rm M_{LN}$), and merger with power-law delay ($\rm M_{PL}$). In Table \ref{tab:diff alphas}, we examine the 1-D KS test results between Parkes/ASKAP subsamples and each of the 4 models by varying the power-law index $\alpha$ 
    between $1.6,$ $1.8,$ and $2.0$. The cutoff energy is fixed at $E_c = 10^{41.5}$~erg. One can see that the $\alpha=1.8$ model can pass all the criteria for both ASKAP and Parkes samples and for all $z$ distribution models. For $\alpha = 2.0$, only the lognormal merger model is rejected by the Parkes FRB sample. All the other $z$ distribution models remain valid. For $\alpha=1.6$, essentially all the $z$-distribution models are rejected. 

\begin{figure*}
\begin{center}
	\includegraphics[width=\columnwidth]{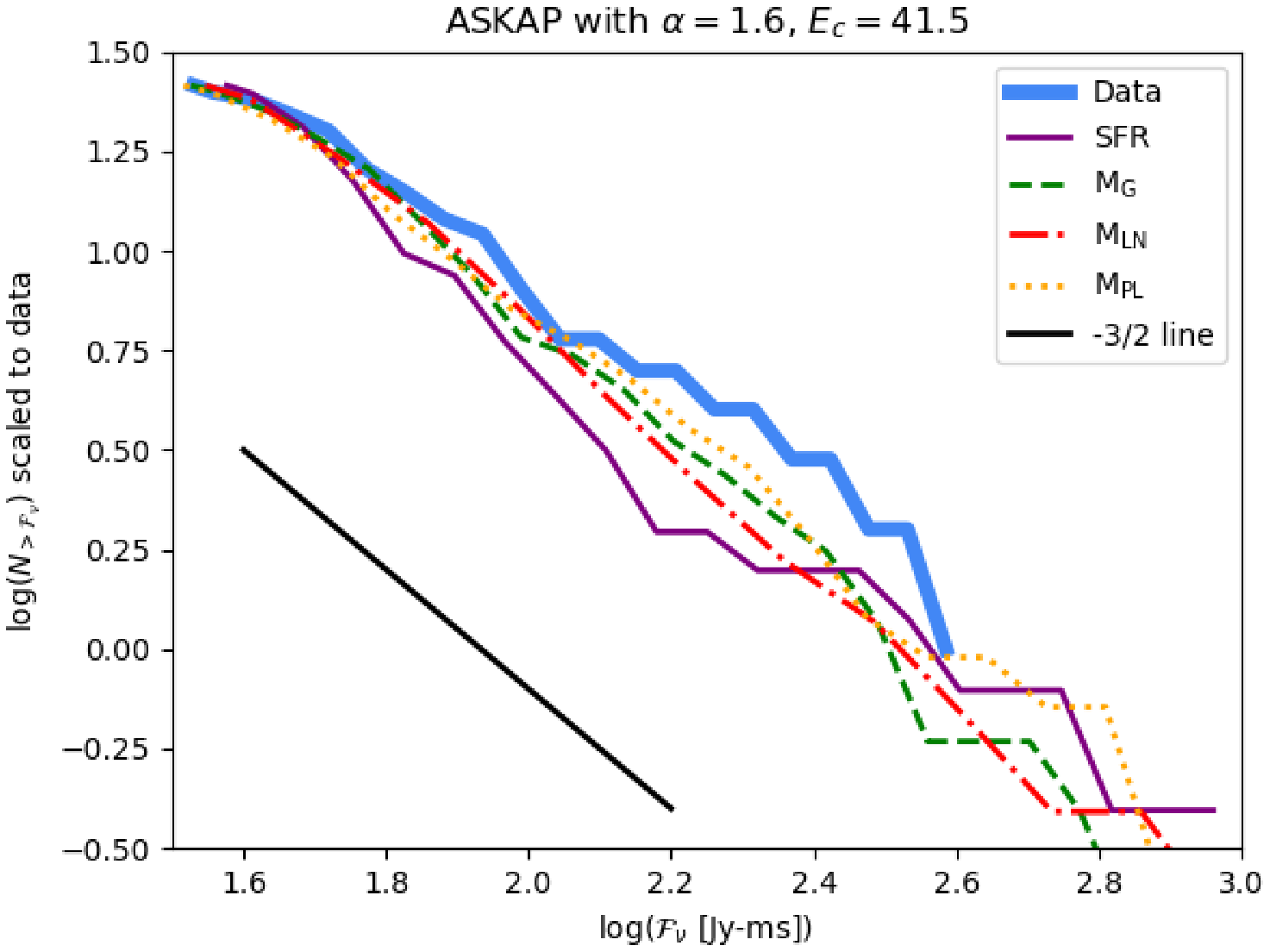}
    \includegraphics[width=\columnwidth]{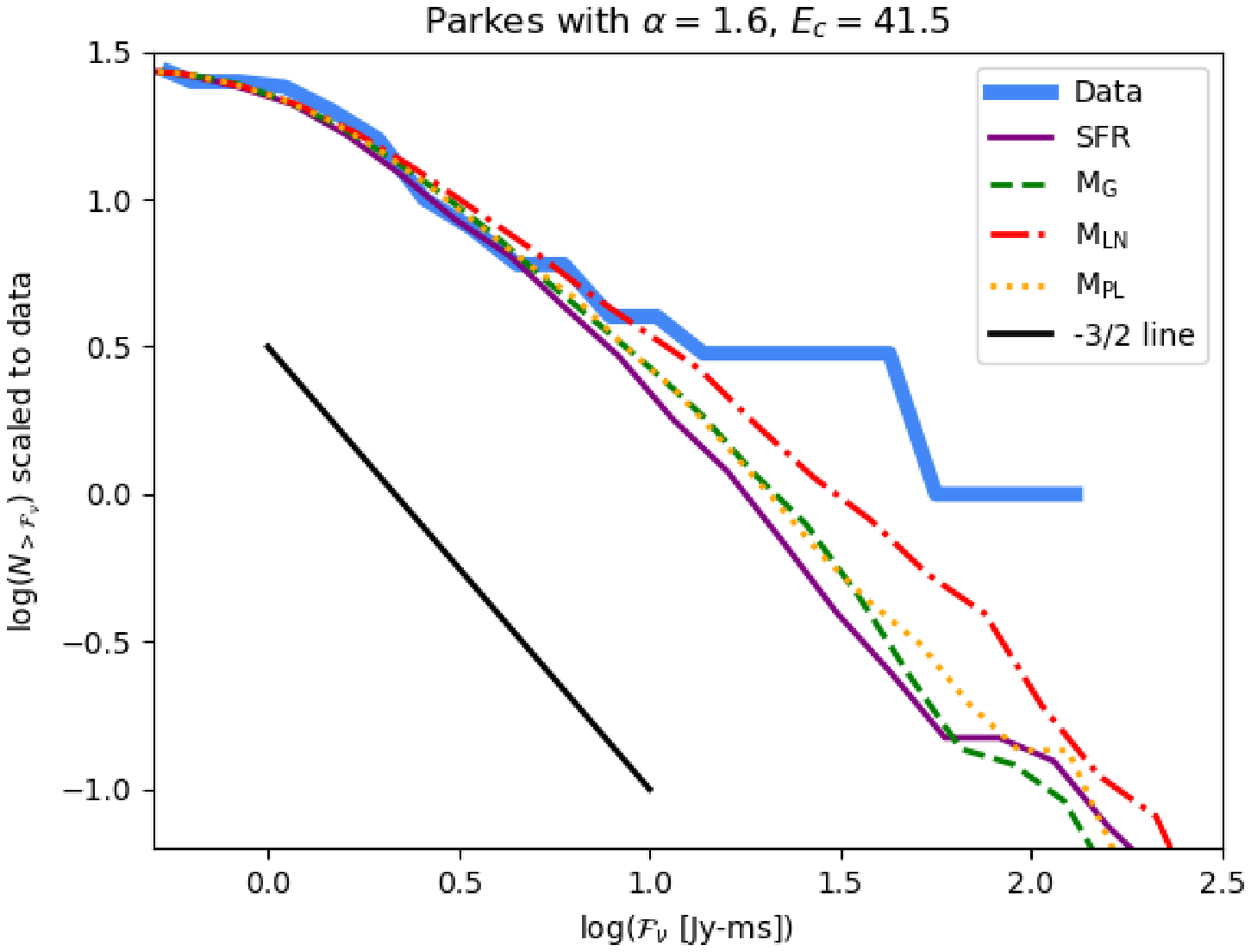}
    \includegraphics[width=\columnwidth]{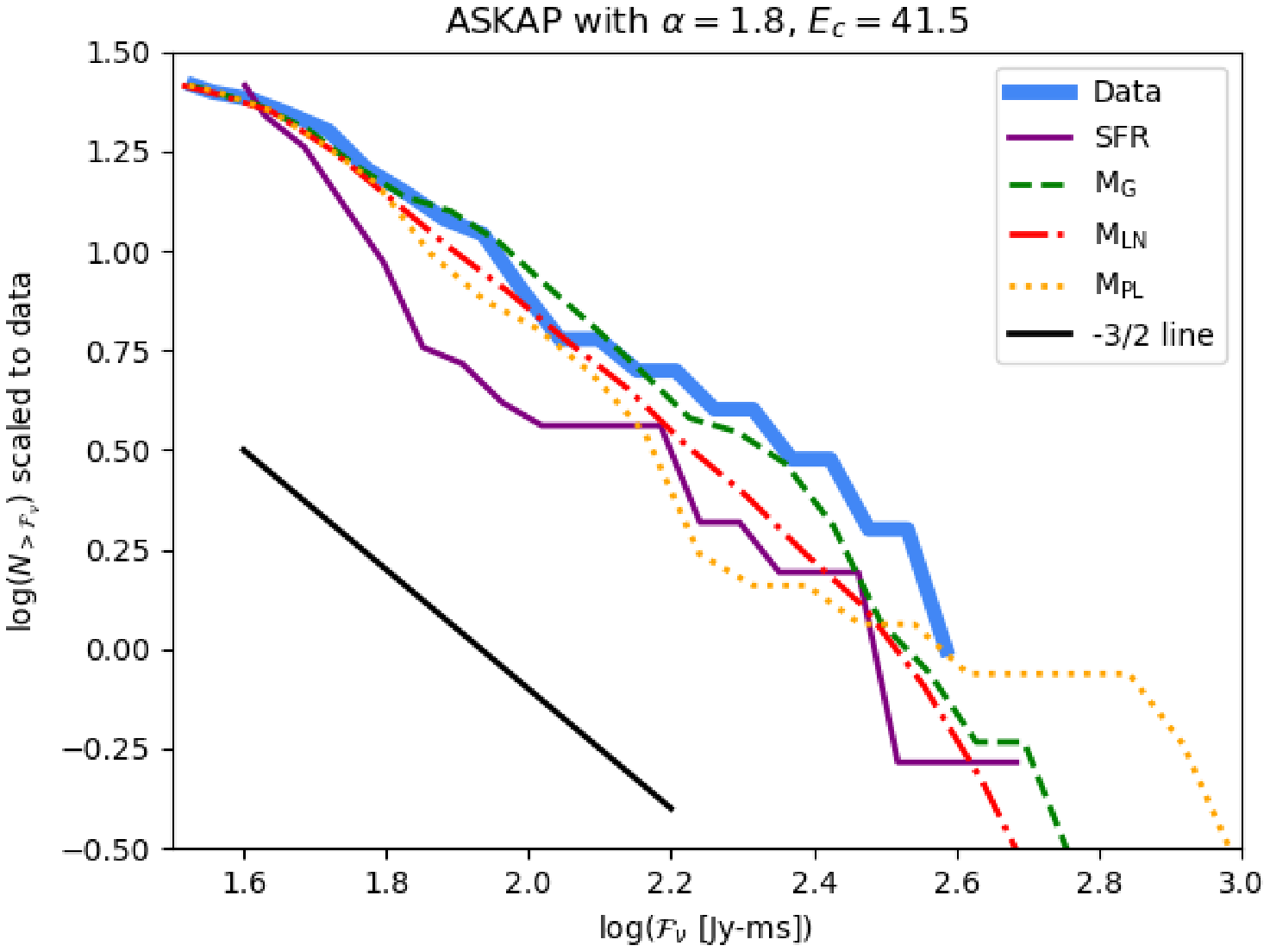}
    \includegraphics[width=\columnwidth]{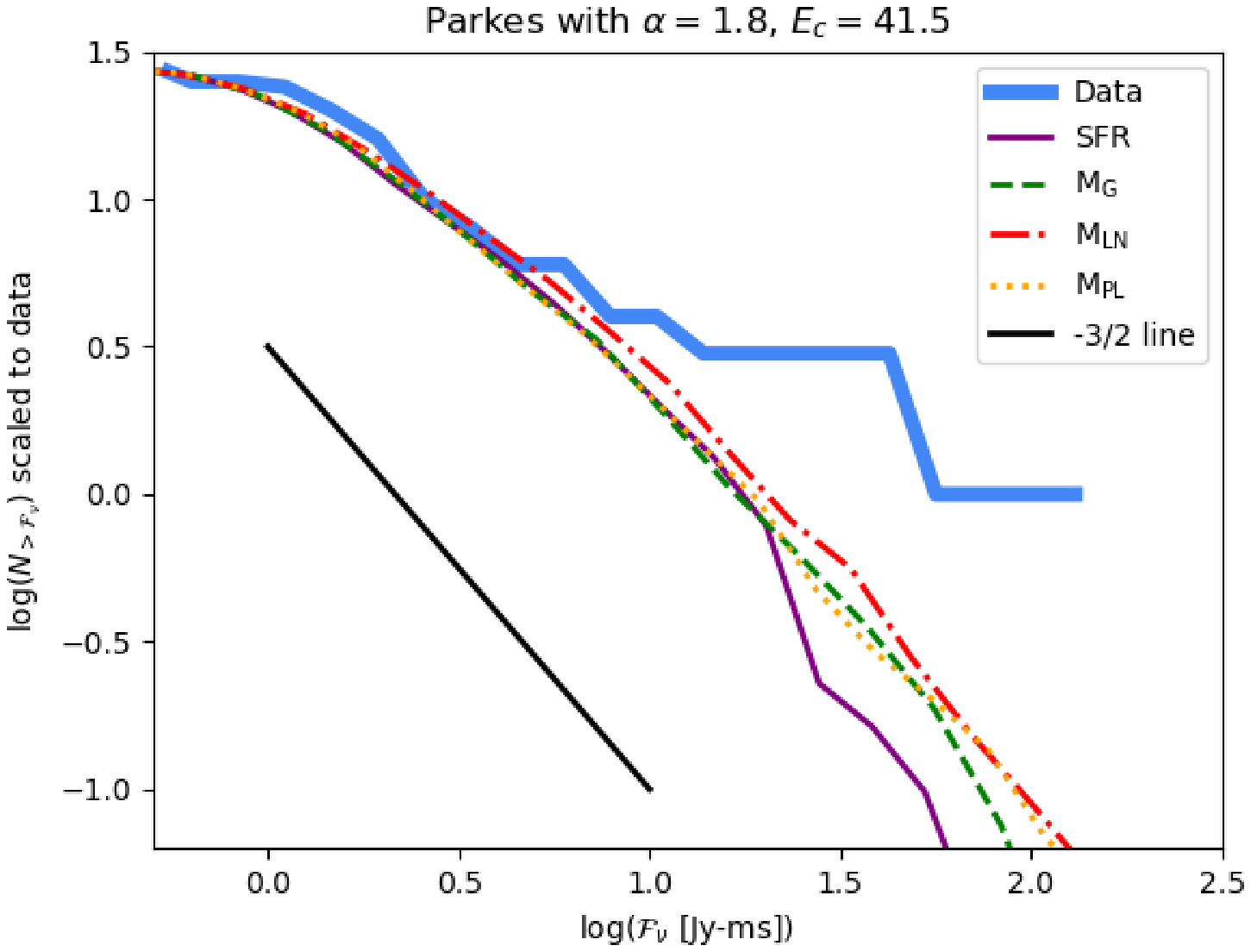}
    \includegraphics[width=\columnwidth]{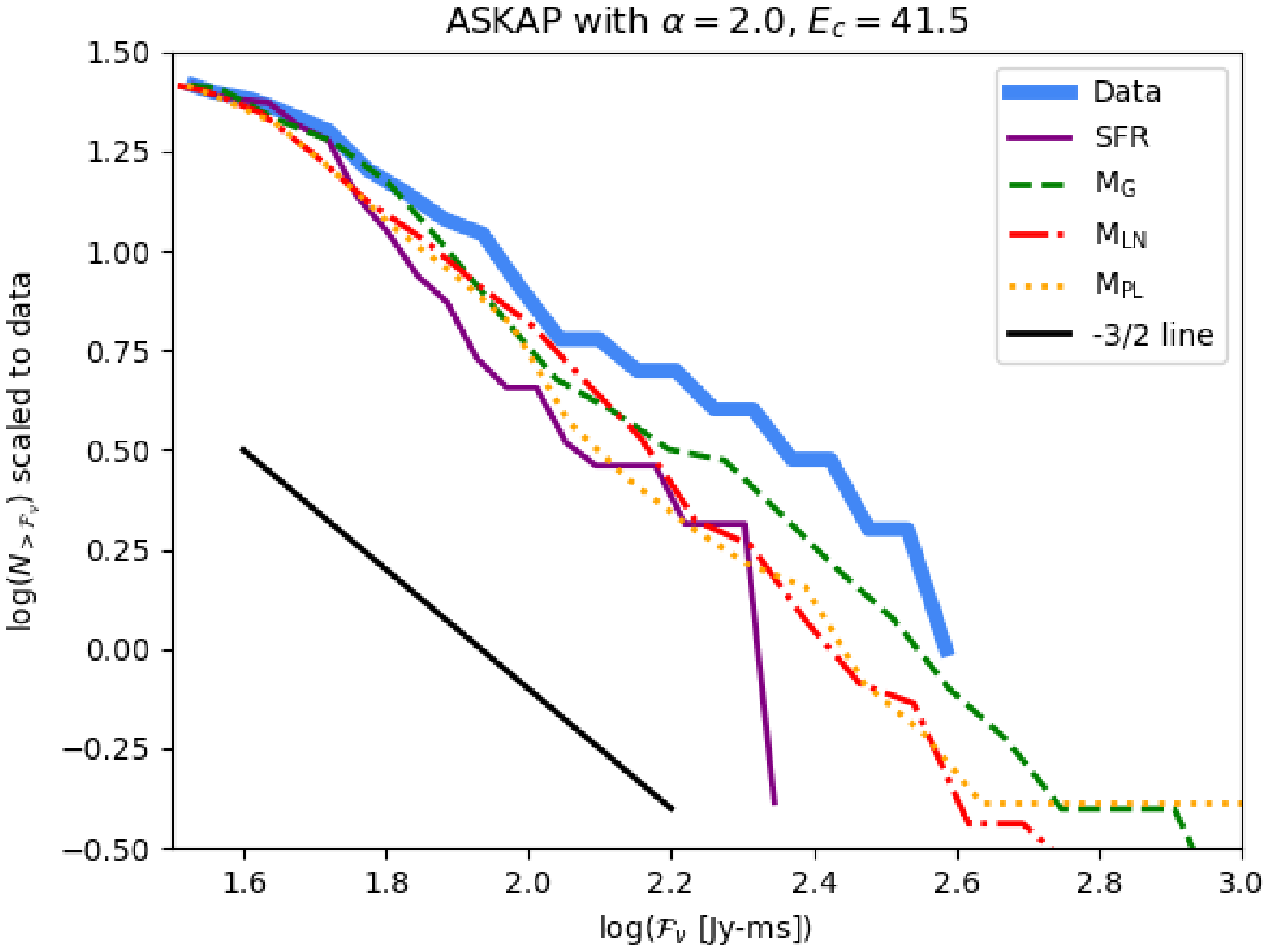}
    \includegraphics[width=\columnwidth]{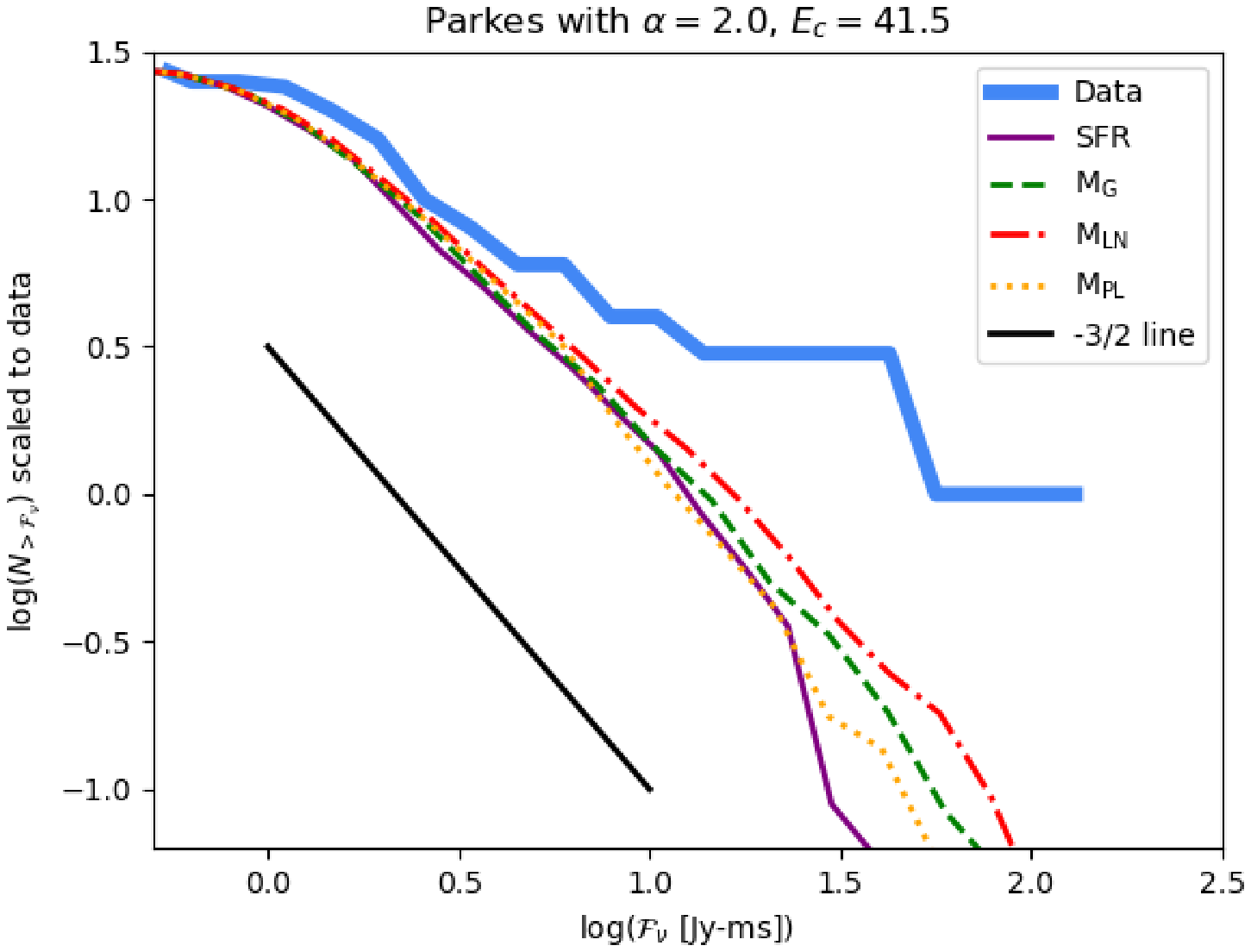}
    \caption{These six panels show the $\log{({\cal F_{\nu}})}$ vs. $\log{(N_{>{\cal F_{\nu}}})}$ plots of the data, SFR (star formation rate history) simulation, $M_{\rm G}$ (Merger with Gaussian delay) simulation, $M_{\rm LN}$ (Merger with lognormal delay) simulation, and $M_{\rm PL}$ (Merger with powerlaw delay) simulation, for three $\alpha$ values of $1.6, 1.8,$ and $2.0$ and two subsamples of the data, the ASKAP and Parkes samples. All of the simulations are scaled to the data. The black solid line with slope $-3/2$ serves as a conventional reference line for Euclidean geometry regardless of energy function.}
    \label{fig:fig3}
\end{center}
\end{figure*}

\begin{figure*}
\begin{center}
    \includegraphics[width=\columnwidth]{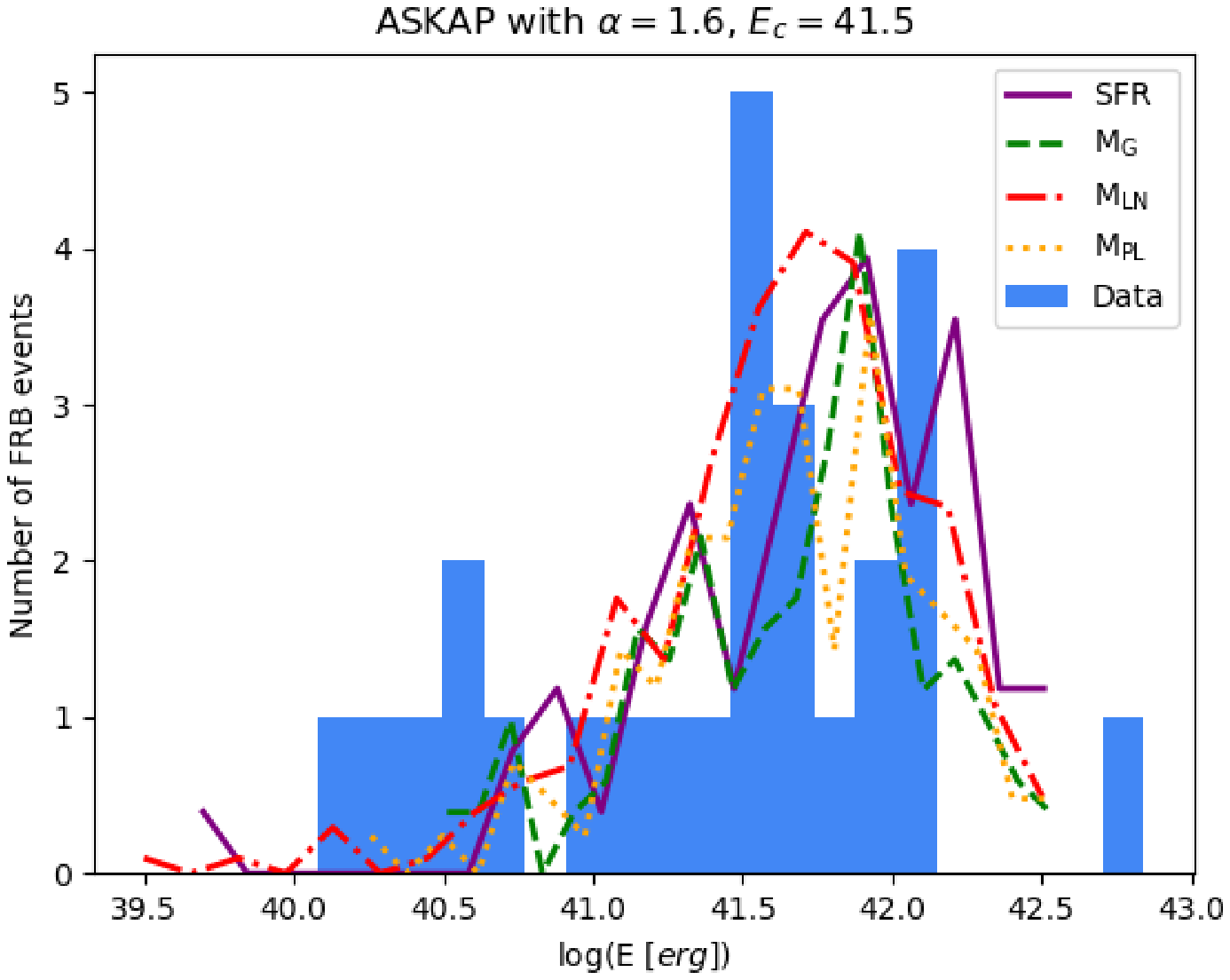}
    \includegraphics[width=\columnwidth]{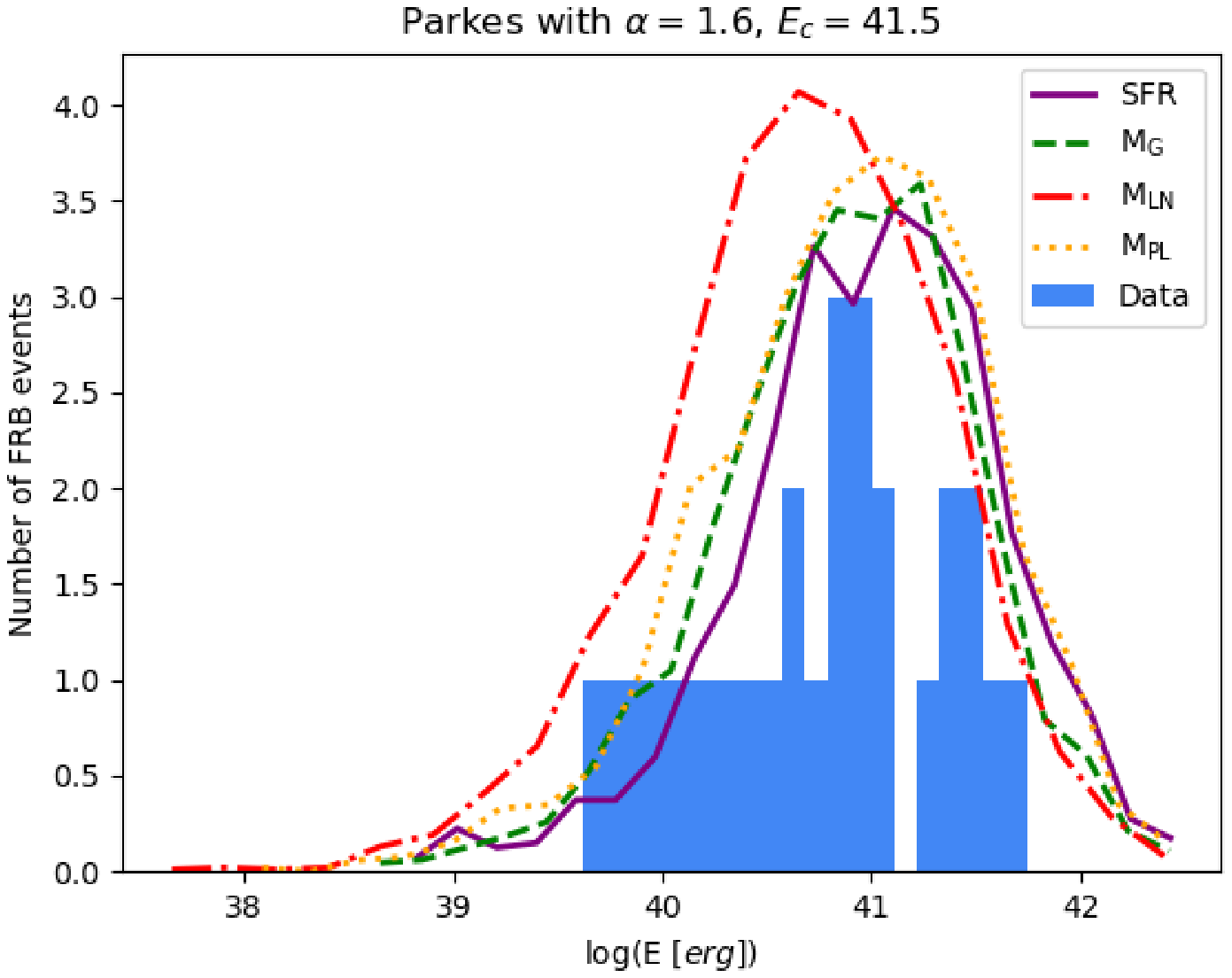}
    \includegraphics[width=\columnwidth]{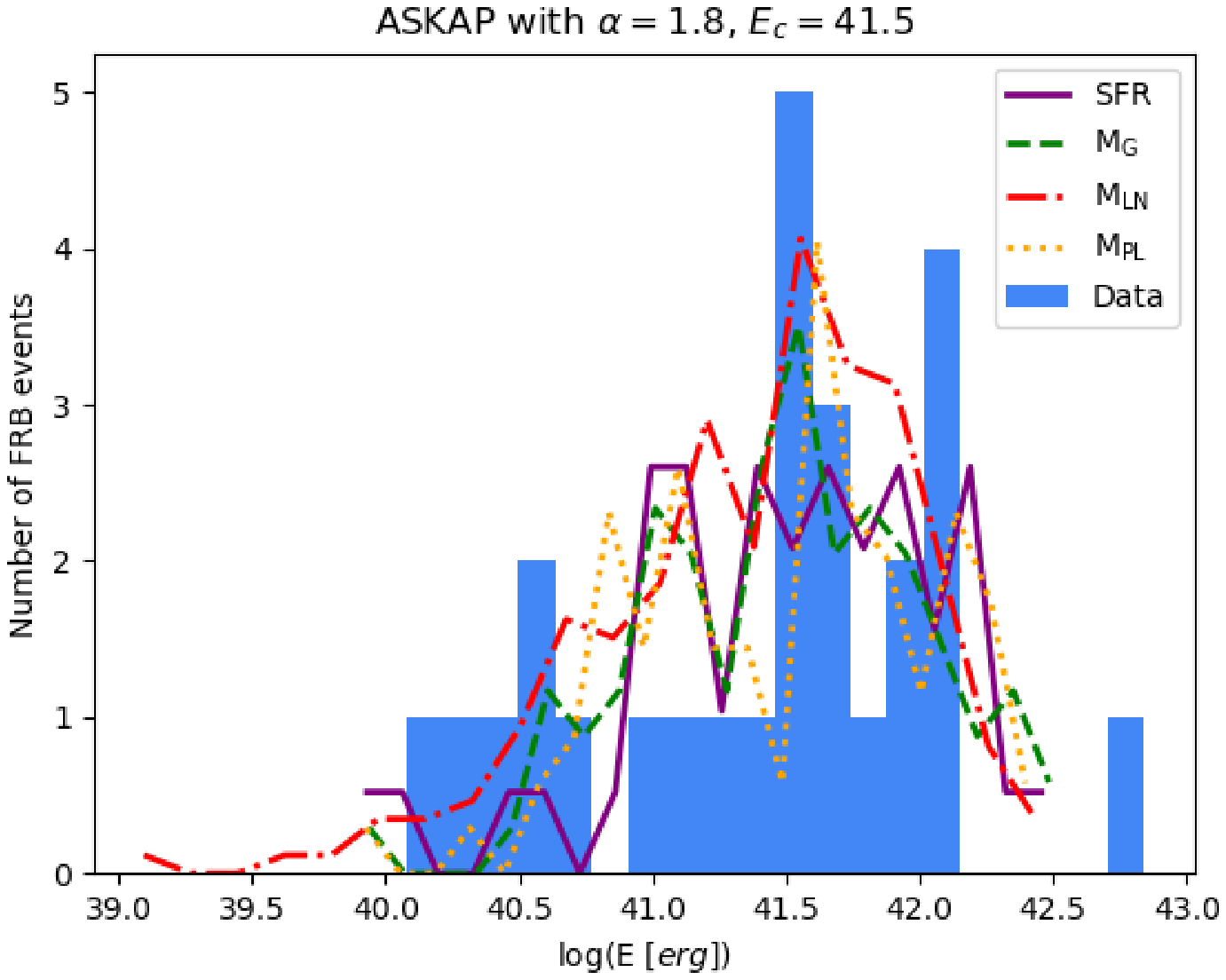}
    \includegraphics[width=\columnwidth]{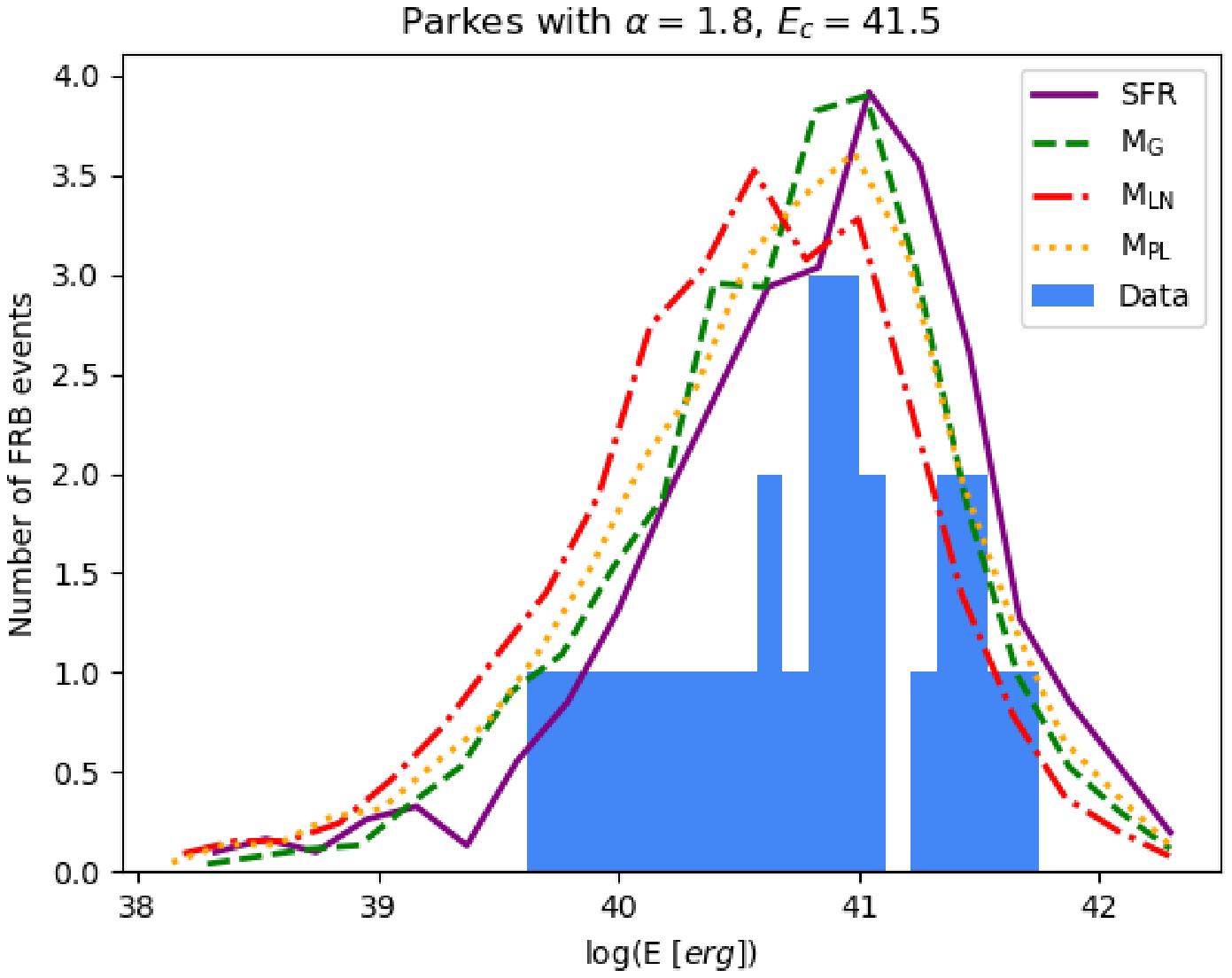}
    \includegraphics[width=\columnwidth]{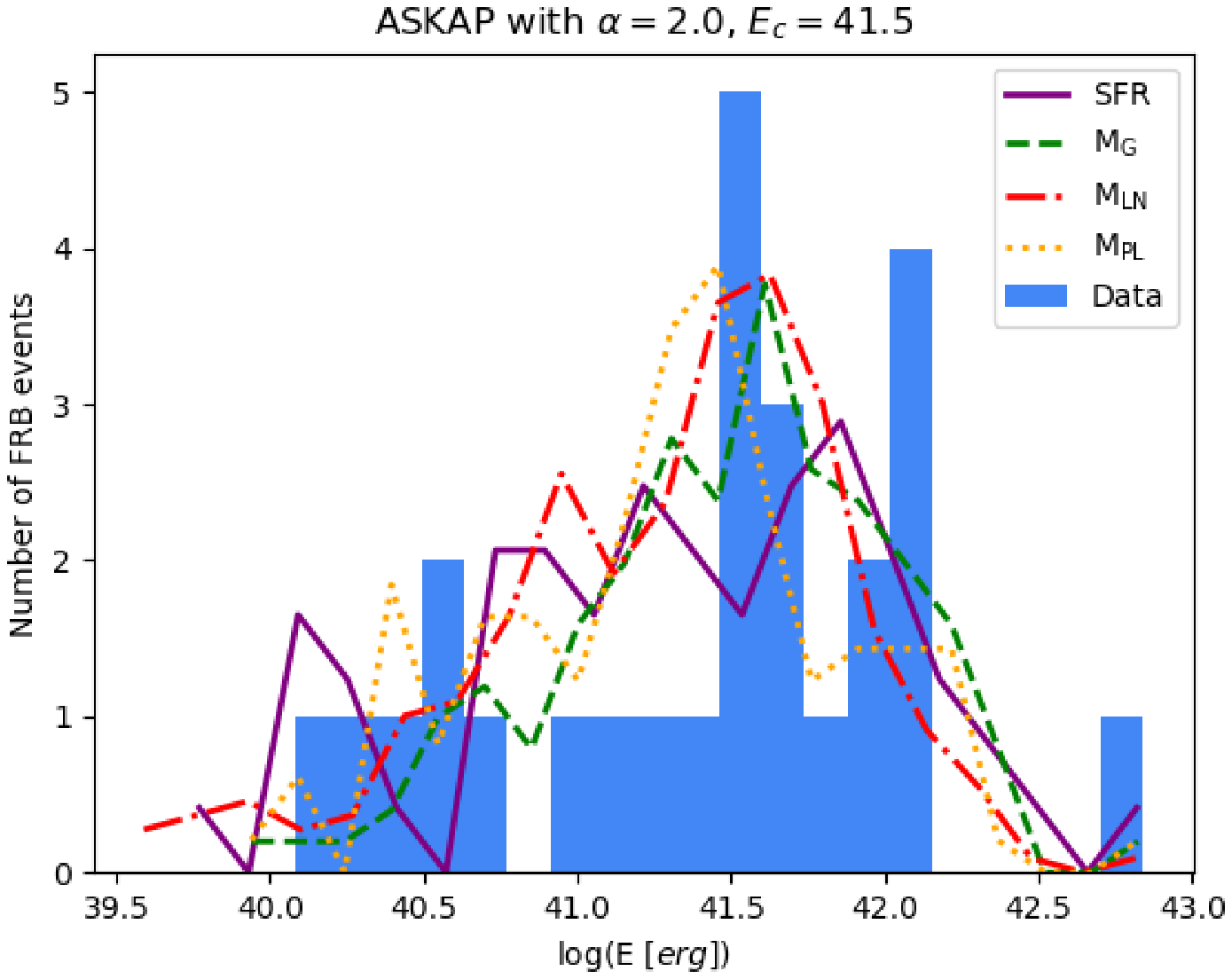}
    \includegraphics[width=\columnwidth]{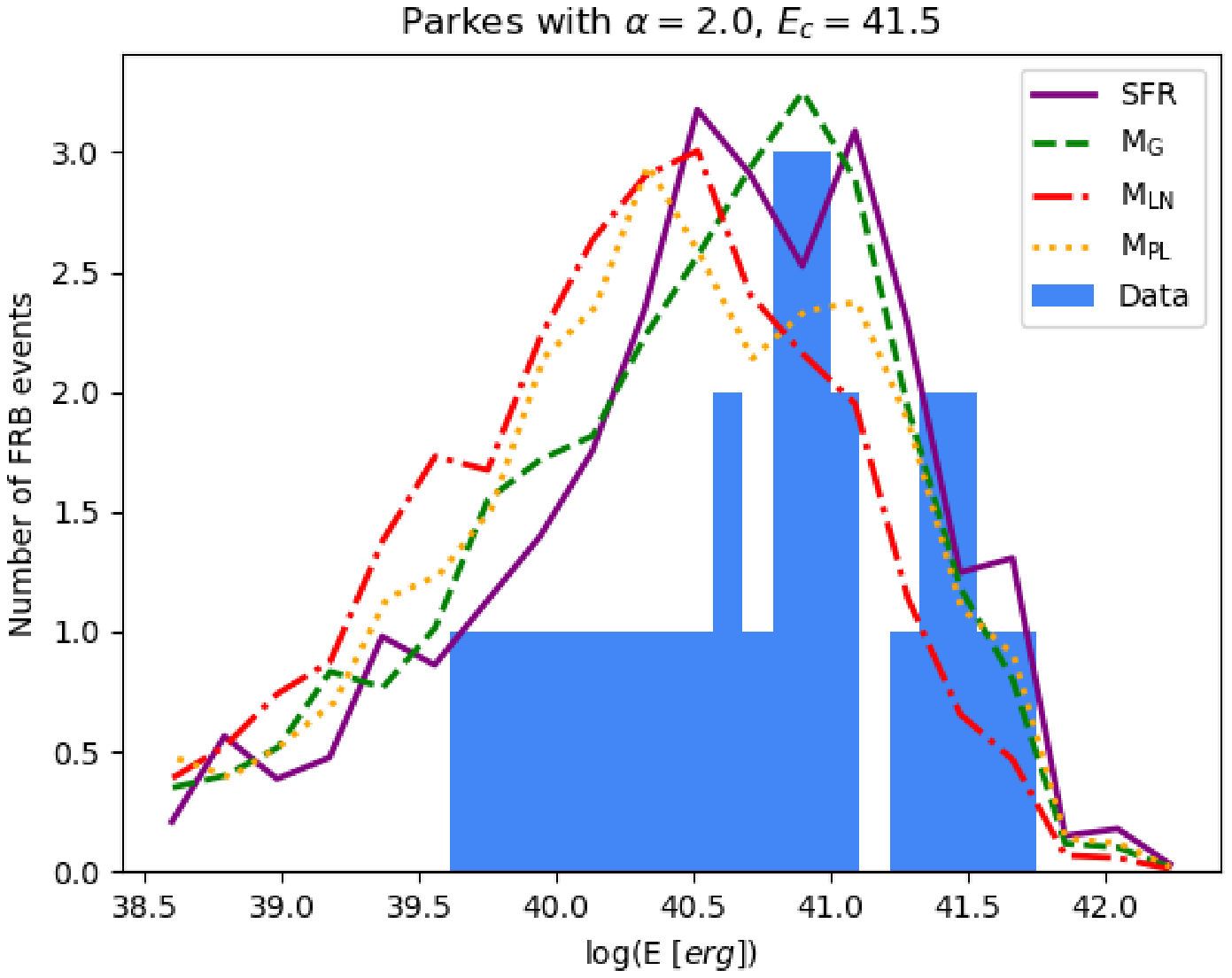}
    \caption{These six panels show the energy distribution plots of the data and four simulation models (notation is consistent with Fig.~\ref{fig:fig3}) for three $\alpha$ values of $1.6,$ $1.8,$ and $2.0$ and two subsamples of the data, the ASKAP and Parkes samples. All of the simulations are scaled to the data.}
    \label{fig:fig4}
\end{center}
\end{figure*}

\begin{figure*}
\begin{center}
    \includegraphics[width=\columnwidth]{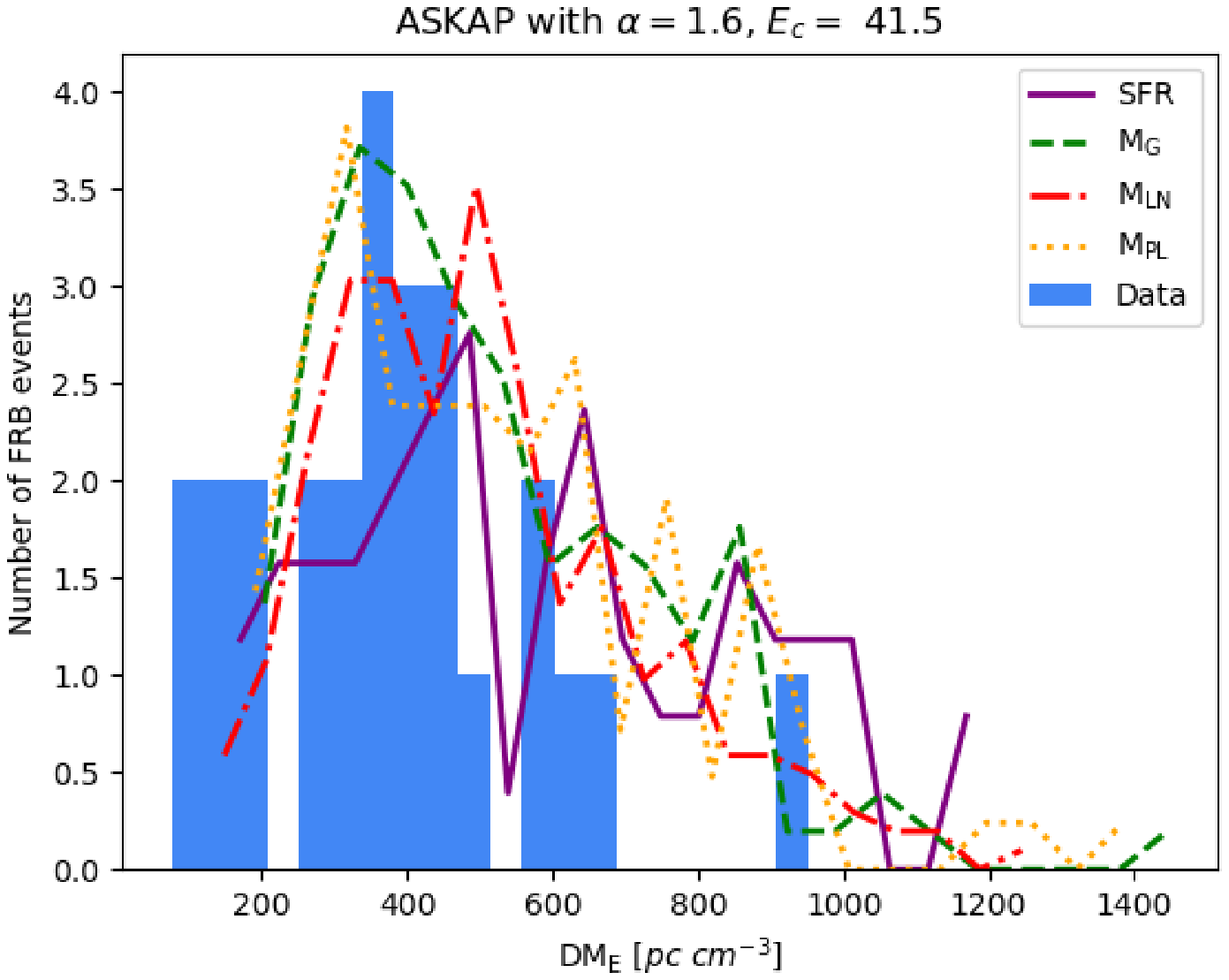}
    \includegraphics[width=\columnwidth]{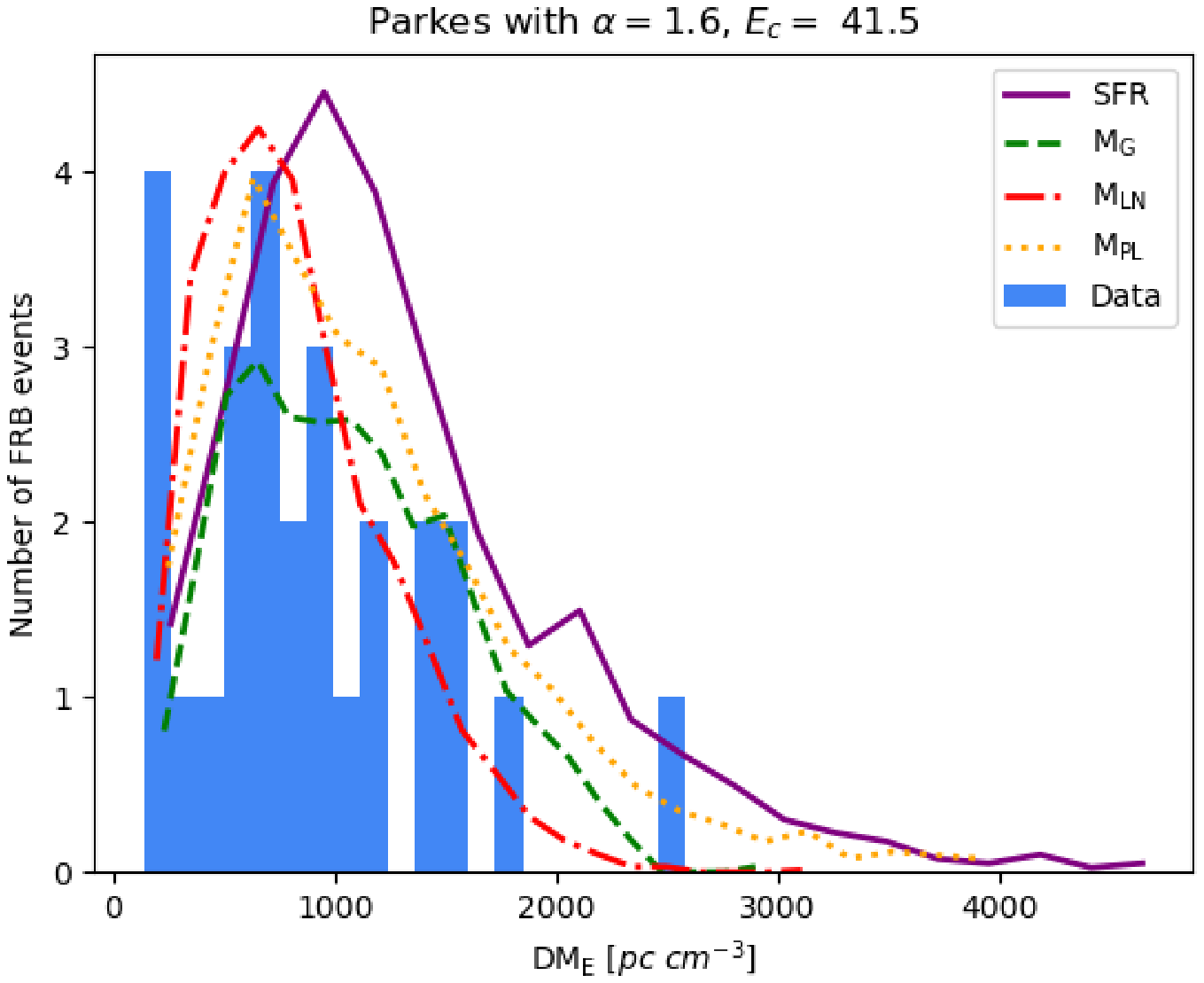}
    \includegraphics[width=\columnwidth]{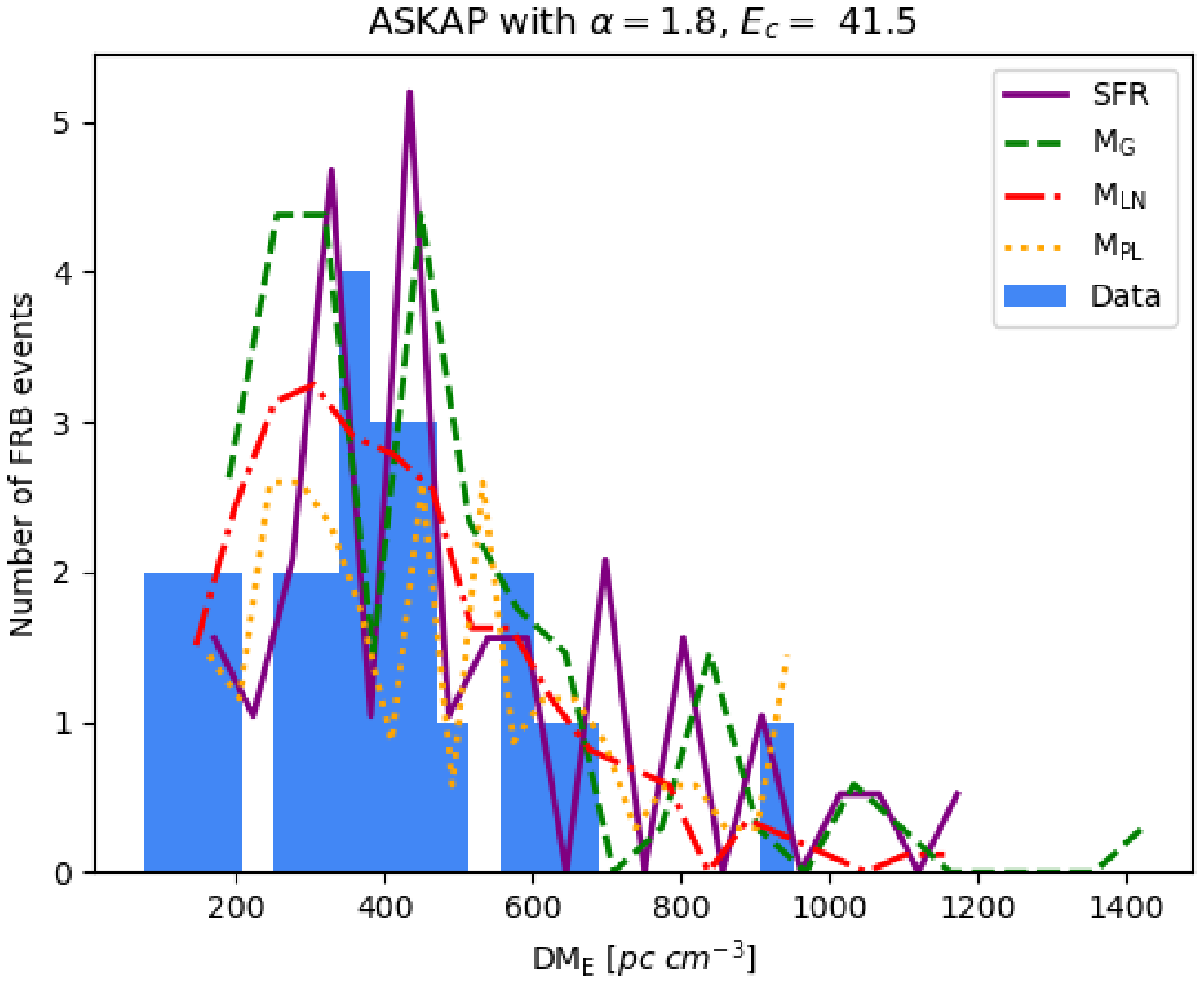}
    \includegraphics[width=\columnwidth]{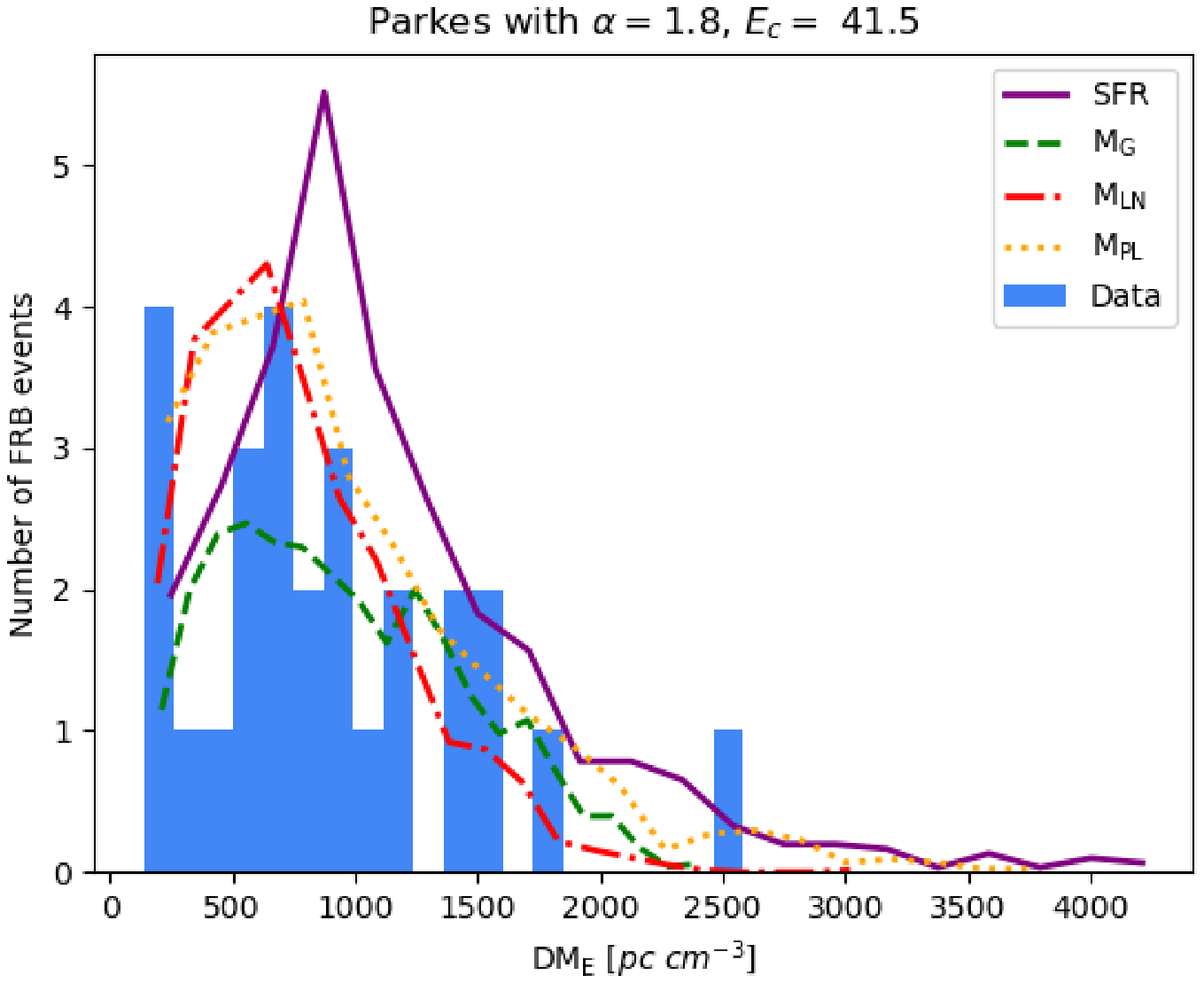}
    \includegraphics[width=\columnwidth]{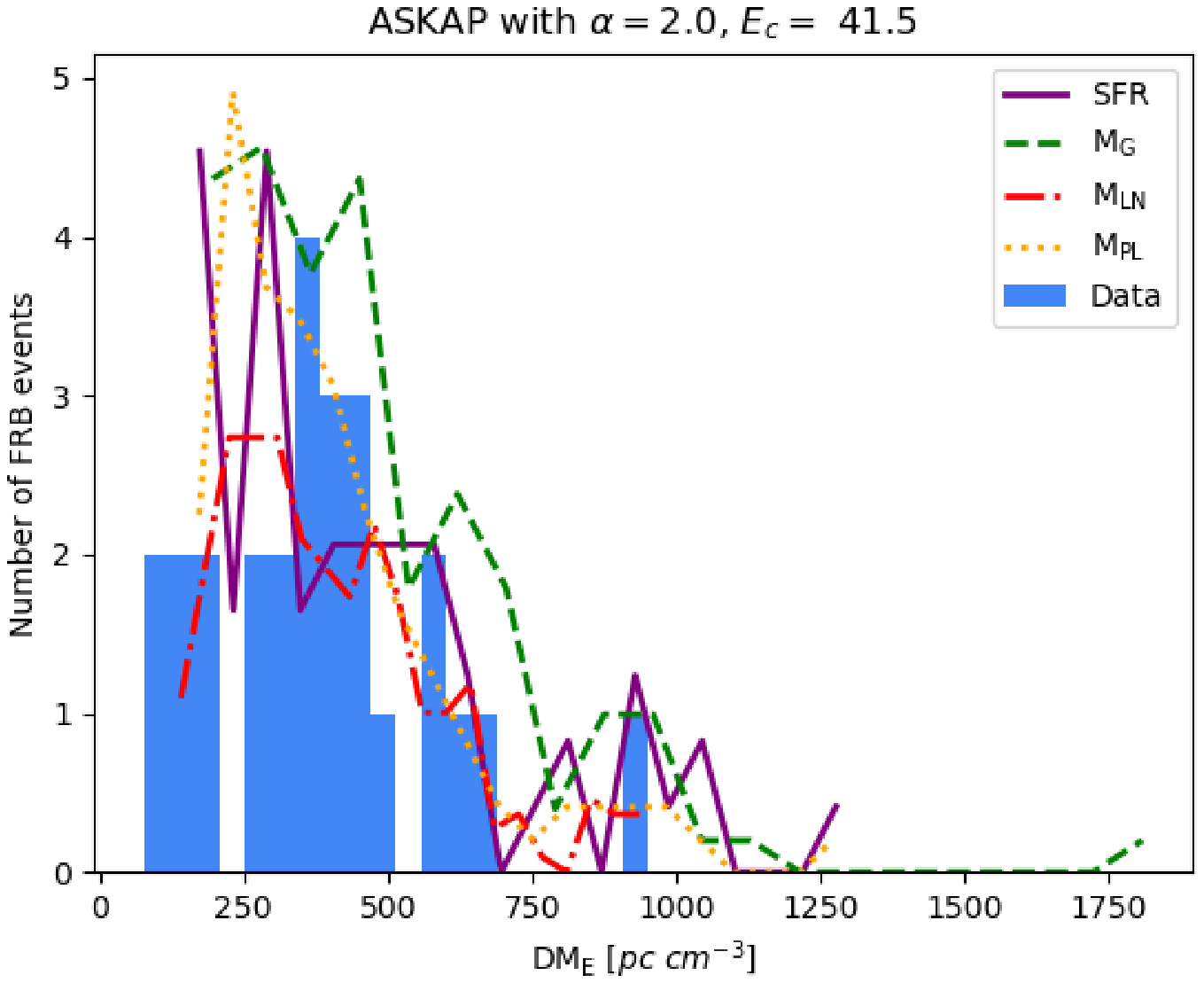}
    \includegraphics[width=\columnwidth]{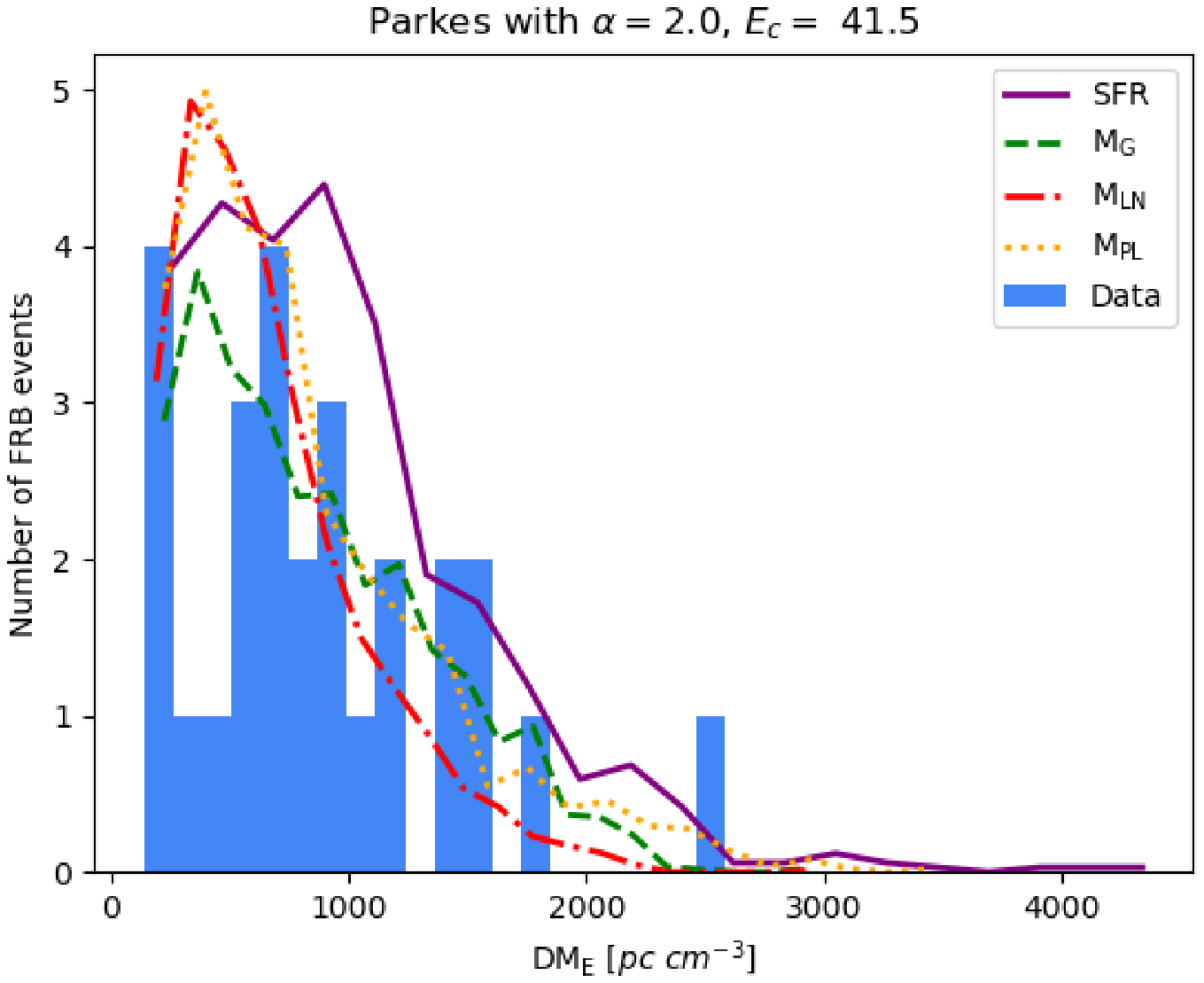}
    \caption{These six panels show the $\rm DM_E$ distribution plots of the data and four simulation models (notation is consistent with Fig.~\ref{fig:fig3}) for three $\alpha$ values of $1.6,$ $1.8,$ and $2.0$ and two subsamples of the data, the ASKAP and Parkes samples. All of the simulations are scaled to the data.}
\end{center}
\label{fig:fig5}
\end{figure*}

    Figs.~\ref{fig:fig3}, \ref{fig:fig4}, and \ref{fig:fig5} show the data vs. model simulations results for $\log{{\cal F}_\nu}$ vs. $\log{N_{>{\cal F}_\nu}}$, energy distribution, and $\rm DM_E$ distribution, respectively. $E_c$ is fixed at $10^{41.5}$~erg, and $\alpha = 1.6,$ $1.8,$ $2.0$, are adopted. These plots give visual impressions on how various models compare with the observational data. The tests on both the ASKAP (left) and Parkes (right) data are presented. For the $\log N_{{\cal F}_\nu} - \log {\cal F}_\nu$ plots (Fig.~\ref{fig:fig3}), the conventional $N_{>{\cal F}_\nu} \propto {\cal F}_\nu^{-3/2}$ line, valid for the Euclidean geometry regardless of the energy function, is plotted in each panel for references. It is seen that the models generally follow this slope in the high-${\cal F}_\nu$ end. The ASKAP sample seems to follow the trend, although the Parkes sample deviates from it at the high end, likely due to small sample statistics and a selection effect in favor of bright events. According to Table \ref{tab:diff alphas}, the $\rm DM_E$ criterion seems to be the most stringent one to eliminate some models. Inspecting Fig.~\ref{fig:fig5}, one can see that for $\alpha=1.6$, there seems to be a deficit of high $\rm DM_E$ data events with respect to predictions, especially for the SFR model. 

    In Table \ref{tab:diff ecut}, we hold $\alpha=1.8$, the best fit power law index, and vary the energy cutoffs between $10^{41.5}, $ $10^{42.5},$ and $10^{43.5}$ erg. We find that all the models pass all the tests, suggesting that both $E_c$ and the redshift distribution are unconstrained for this value of $\alpha$.

\begin{table*}
    \centering
    \begin{tabular}{c|ccc|ccc|ccc|ccc}
        \hline
    $\log(E_c)=41.5$   & \multicolumn{3}{c}{SFR}
        &
        \multicolumn{3}{c}{Merger (G)}
        &
        \multicolumn{3}{c}{Merger (LN)} 
        &
        \multicolumn{3}{c}{Merger (PL)} \\\hline
        {$\alpha$}  & 1.6 & 1.8 & 2.0 & 1.6 & 1.8 & 2.0 & 1.6 & 1.8 & 2.0 & 1.6 & 1.8 & 2.0 \\
        \hline
        \multicolumn{13}{c}{Specific Fluence}   \\
        \hline
        {ASKAP} &  $\checkmark$ & $\checkmark$ & $\checkmark$ & $\checkmark$ & $\checkmark$ & $\checkmark$ & $\checkmark$ & $\checkmark$ & $\checkmark$ & $\checkmark$ & $\checkmark$ & $\checkmark$\\
        {Parkes} & $\checkmark$ & $\checkmark$ & $\checkmark$ & $\checkmark$ & $\checkmark$ & $\checkmark$ & $\checkmark$ & $\checkmark$ & $\checkmark$ & $\checkmark$ & $\checkmark$ & $\checkmark$\\
        \hline 
        \multicolumn{13}{c}{Energy}   \\
        \hline 
        {ASKAP} & $\checkmark$ & $\checkmark$ & $\checkmark$ & $\checkmark$ & $\checkmark$ & $\checkmark$ & $\checkmark$ & $\checkmark$ & $\checkmark$ & $\checkmark$ & $\checkmark$ & $\checkmark$\\
        {Parkes} & $\checkmark$ & $\checkmark$ & $\checkmark$ & $\checkmark$ & $\checkmark$ & $\checkmark$ & $\checkmark$ & $\checkmark$ & X & $\checkmark$ & $\checkmark$ & $\checkmark$\\
        \hline
        \multicolumn{13}{c}{$\rm DM_E$}    \\
        \hline
        {ASKAP} & X & $\checkmark$ & $\checkmark$ & X & $\checkmark$ & $\checkmark$ & X & $\checkmark$ & $\checkmark$ & X & $\checkmark$ & $\checkmark$\\
        {Parkes} & X & $\checkmark$ & $\checkmark$ & $\checkmark$ & $\checkmark$ & $\checkmark$ & $\checkmark$ & $\checkmark$ & X & $\checkmark$ & $\checkmark$ & $\checkmark$\\
        \hline
        \multicolumn{13}{c}{$2$D Energy and $\rm DM_E$} \\
        \hline
        {ASKAP} & X & $\checkmark$ & $\checkmark$ & $\checkmark$ & $\checkmark$ & $\checkmark$ & $\checkmark$ & $\checkmark$ & $\checkmark$ & $\checkmark$  & $\checkmark$ & $\checkmark$\\
        {Parkes} & $\checkmark$  & $\checkmark$ & $\checkmark$ & $\checkmark$ & $\checkmark$ & $\checkmark$ & $\checkmark$ & $\checkmark$ & X & $\checkmark$ & $\checkmark$ & $\checkmark$\\
        \hline
        
    \end{tabular}
    \caption{KS test results for each of the four models compared with the data of the ASKAP and Parkes subsamples. We hold the energy cutoff constant, where $\log(E_c) = 41.5$, and we consider three powerlaw indices for the energy distribution of the simulations (1.6, 1.8, 2.0). The entries labeled "X" refer to the null hypothesis that the data and simulation come from the same distribution being rejected (KS statistic > critical value), and "$\checkmark$" refers to the null hypothesis not being rejected (KS statistic $\leq$ critical value).}
    \label{tab:diff alphas}
\end{table*}
    
\begin{table*}
    \centering
    \begin{tabular}{c|ccc|ccc|ccc|ccc}
        \hline
    $\alpha = 1.8$   & \multicolumn{3}{c}{SFR}
        &
        \multicolumn{3}{c}{Merger (G)}
        &
        \multicolumn{3}{c}{Merger (LN)} 
        &
        \multicolumn{3}{c}{Merger (PL)} \\\hline
        {$\log{E_{cut}}$}  & 41.5 & 42.5 & 43.5 & 41.5 & 42.5 & 43.5 & 41.5 & 42.5 & 43.5 & 41.5 & 42.5 & 43.5 \\
        \hline
        \multicolumn{13}{c}{Specific Fluence}   \\
        \hline
        {ASKAP} & $\checkmark$ & $\checkmark$ & $\checkmark$ & $\checkmark$ & $\checkmark$ & $\checkmark$ & $\checkmark$ & $\checkmark$ & $\checkmark$ & $\checkmark$ & $\checkmark$ & $\checkmark$\\
        {Parkes} & $\checkmark$ & $\checkmark$ & $\checkmark$ & $\checkmark$ & $\checkmark$ & $\checkmark$ & $\checkmark$ & $\checkmark$ & $\checkmark$ & $\checkmark$ & $\checkmark$ & $\checkmark$\\
        \hline 
        \multicolumn{13}{c}{Energy}   \\
        \hline 
        {ASKAP} & $\checkmark$ & $\checkmark$ & $\checkmark$ & $\checkmark$ & $\checkmark$ & $\checkmark$ & $\checkmark$ & $\checkmark$ & $\checkmark$ & $\checkmark$ & $\checkmark$ & $\checkmark$\\
        {Parkes} & $\checkmark$ & $\checkmark$ & $\checkmark$ & $\checkmark$ & $\checkmark$ & $\checkmark$ & $\checkmark$ & $\checkmark$ & $\checkmark$ & $\checkmark$ & $\checkmark$ & $\checkmark$\\
        \hline
        \multicolumn{13}{c}{$\rm DM_E$}    \\
        \hline
        {ASKAP} & $\checkmark$ & $\checkmark$ & $\checkmark$ & $\checkmark$ & $\checkmark$ & $\checkmark$ & $\checkmark$ & $\checkmark$ & $\checkmark$ & $\checkmark$ & $\checkmark$ & $\checkmark$\\
        {Parkes} & $\checkmark$ & $\checkmark$ & $\checkmark$ & $\checkmark$ & $\checkmark$ & $\checkmark$ & $\checkmark$ & $\checkmark$ & $\checkmark$ & $\checkmark$ & $\checkmark$ & $\checkmark$\\
        \hline
        \multicolumn{13}{c}{$2$D Energy and $\rm DM_E$} \\
        \hline
        {ASKAP} & $\checkmark$ & $\checkmark$ & $\checkmark$ & $\checkmark$ & $\checkmark$ & $\checkmark$ & $\checkmark$ & $\checkmark$ & $\checkmark$ & $\checkmark$  & $\checkmark$ & $\checkmark$\\
        {Parkes} & $\checkmark$  & $\checkmark$ & $\checkmark$ & $\checkmark$ & $\checkmark$ & $\checkmark$ & $\checkmark$ & $\checkmark$ & $\checkmark$ & $\checkmark$ & $\checkmark$ & $\checkmark$\\
        \hline
    \end{tabular}
    \caption{KS test results for each of the four models compared with the data of the ASKAP and Parkes subsamples. We hold the powerlaw index constant, where $\alpha = 1.8$, and we consider three energy cutoffs for the energy distribution of the simulations ($\log{E_{\rm cut}} = 41.5, 42.5, 43.5$). The entries labeled "X" refer to the null hypothesis that the data and simulation come from the same distribution being rejected (KS statistic > critical value), and "$\checkmark$" refers to the null hypothesis not being rejected (KS statistic $\leq$ critical value).}
    \label{tab:diff ecut}
\end{table*}

\subsection{Two-Dimensional KS Test on $\rm \bf DM_E$ and Energy}
\label{KS2D}

    \subsubsection{Method}
    Since the observed distributions of energy and $\rm DM_E$ carry independent information regarding the intrinsic energy and redshift distributions of FRBs, it is worth investigating a 2-D KS test for all the models against the two FRB sub-samples. 2-D KS tests can be sometimes more constraining than individual 1-D KS tests.
    
    When generalizing the classical 1-D KS test to higher dimensions, one needs to consider the maximum difference between two samples $\rm D_{KS}$ when all possible directions along each axis is considered. \cite{fasano87} proposed an efficient way of calculating such a result. For two dimensions, $\rm D_{KS}$ is calculated in four quadrants of $n$ origins for both samples individually:
    \begin{eqnarray} 
        \left\{ 
          \begin{array}{l}
             (x<X_i, y<Y_i)    \\
             (x<X_i, y>Y_i)    \\
             (x>X_i, y<Y_i)    \\
             (x>X_i, y>Y_i)
          \end{array}
        \right. , 
        & (i=1, ..., n).
    \end{eqnarray}
    Then, according to \cite{fasano87}, we take the average $\bar{D}_{\rm KS}$ of the two $D_{\rm KS}$ results from each sample. 
    
    One requirement to use the two-sample test from \cite{fasano87} that is not necessarily true in our case is that the probability distribution should be easily recovered from that of one-sample tests. We resolve this by estimating the null probability distribution using Monte Carlo simulations. In each trial of the simulation, we randomly select a data-sized sample of the model simulation sample and calculate the $D_{\rm KS}$ result. After running 1000 such simulations, we are able to get the null probability distribution of $D_{\rm KS}$. Thus, for each iteration of our 2-D KS value calculation, we interpolate the $D_{\rm KS}$ in the simulated $D_{\rm KS}$ probability distribution. 
    \subsubsection{Results}
\begin{figure*}
	\includegraphics[width=\columnwidth]{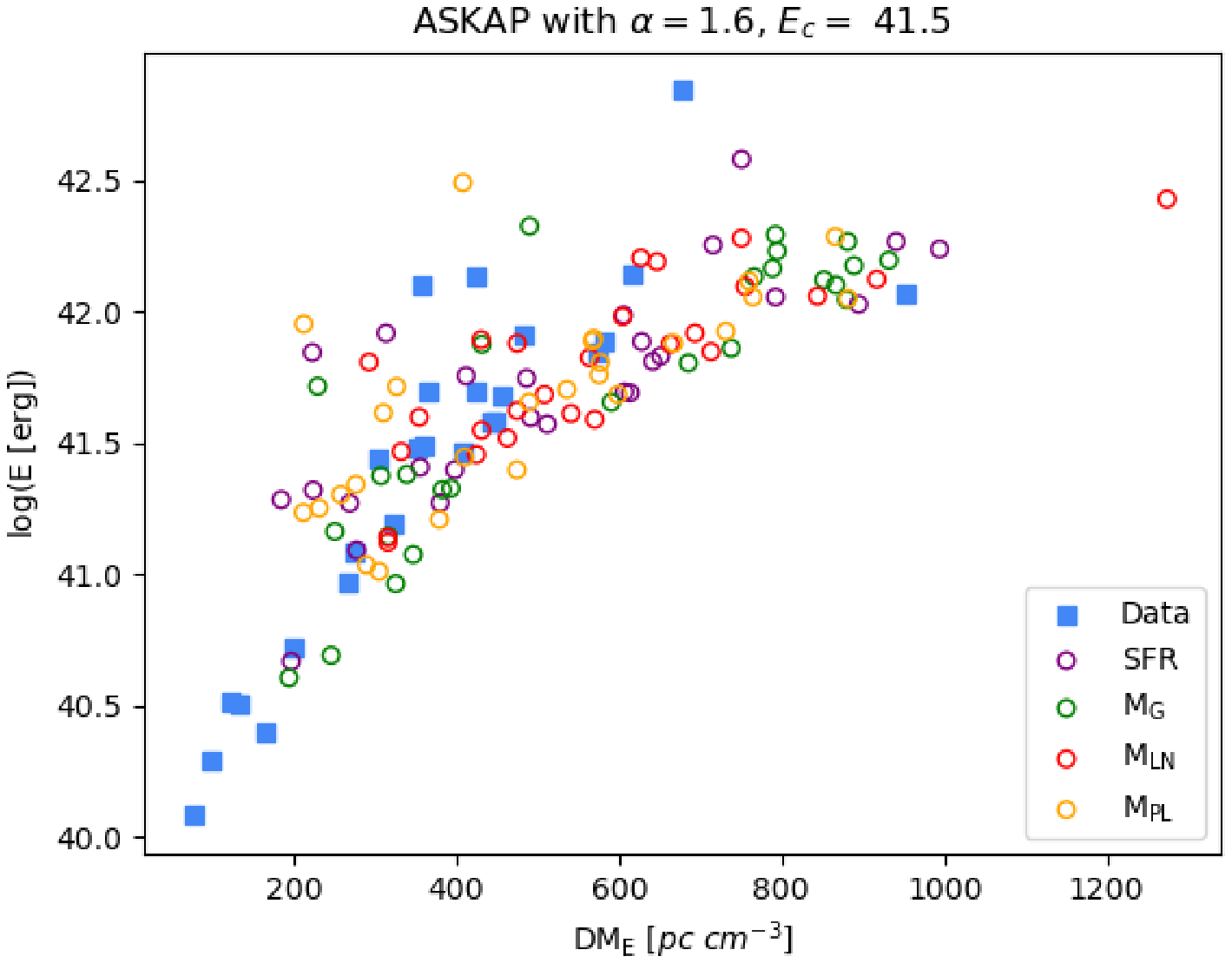}
    \includegraphics[width=\columnwidth]{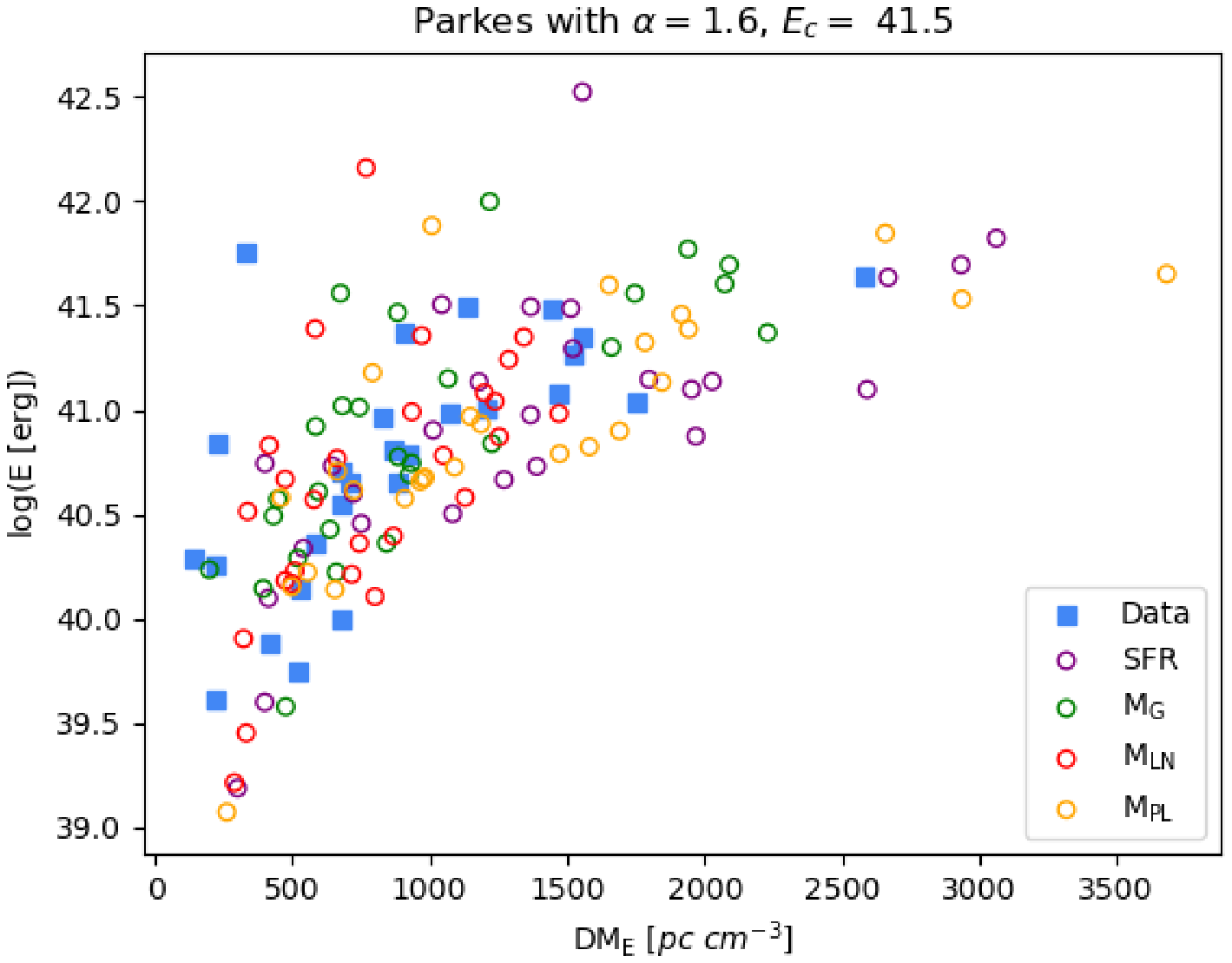}
    \includegraphics[width=\columnwidth]{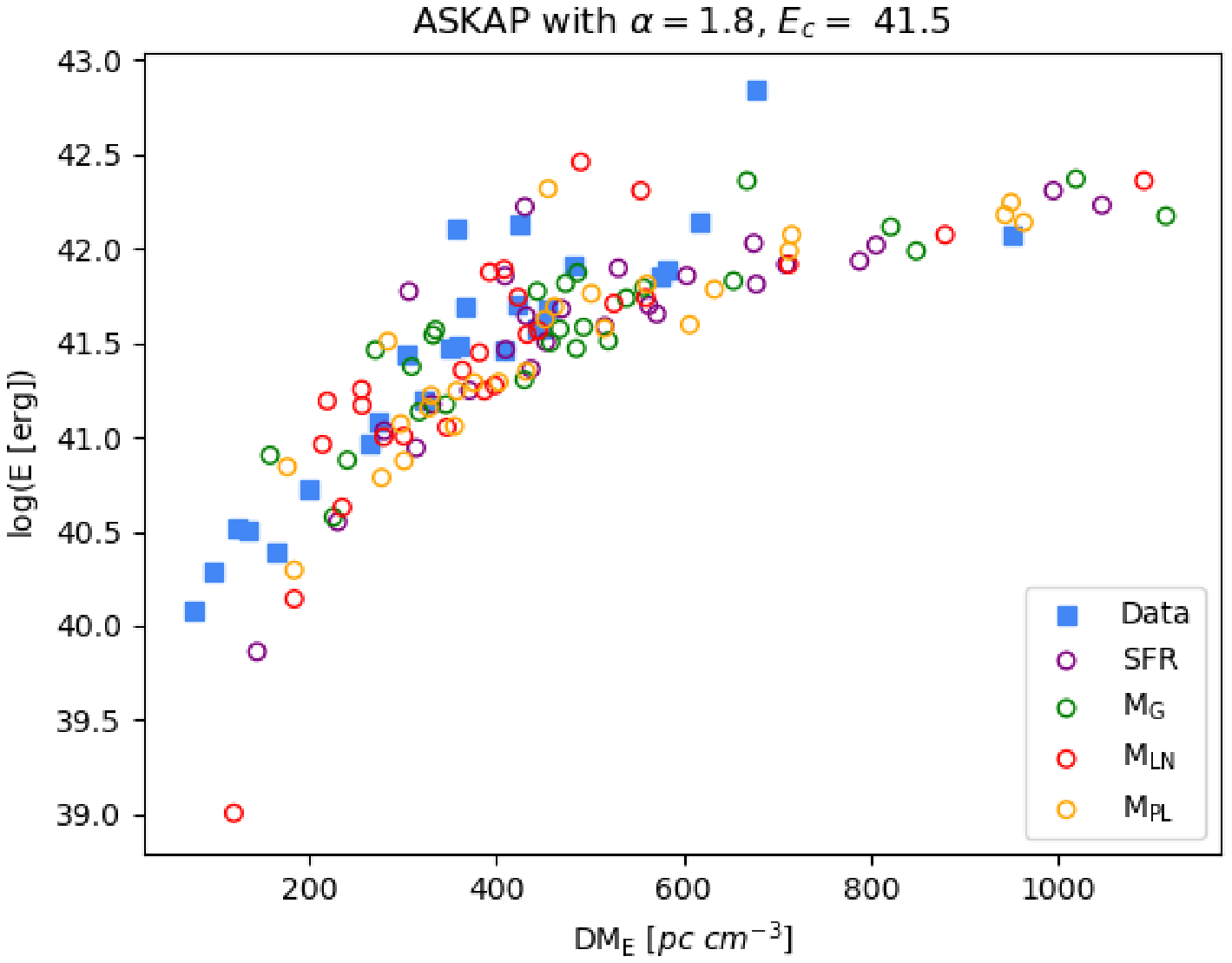}
    \includegraphics[width=\columnwidth]{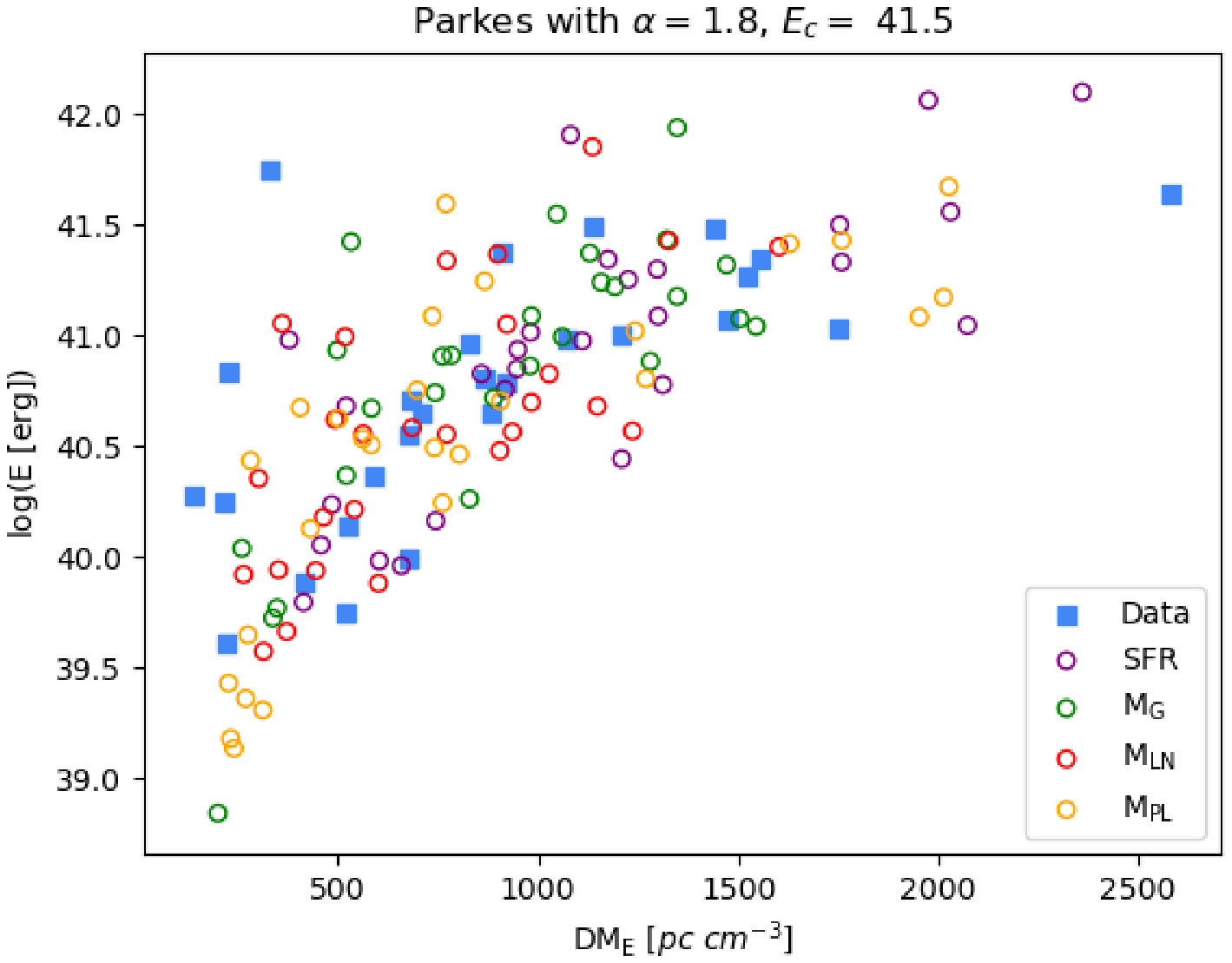}
    \includegraphics[width=\columnwidth]{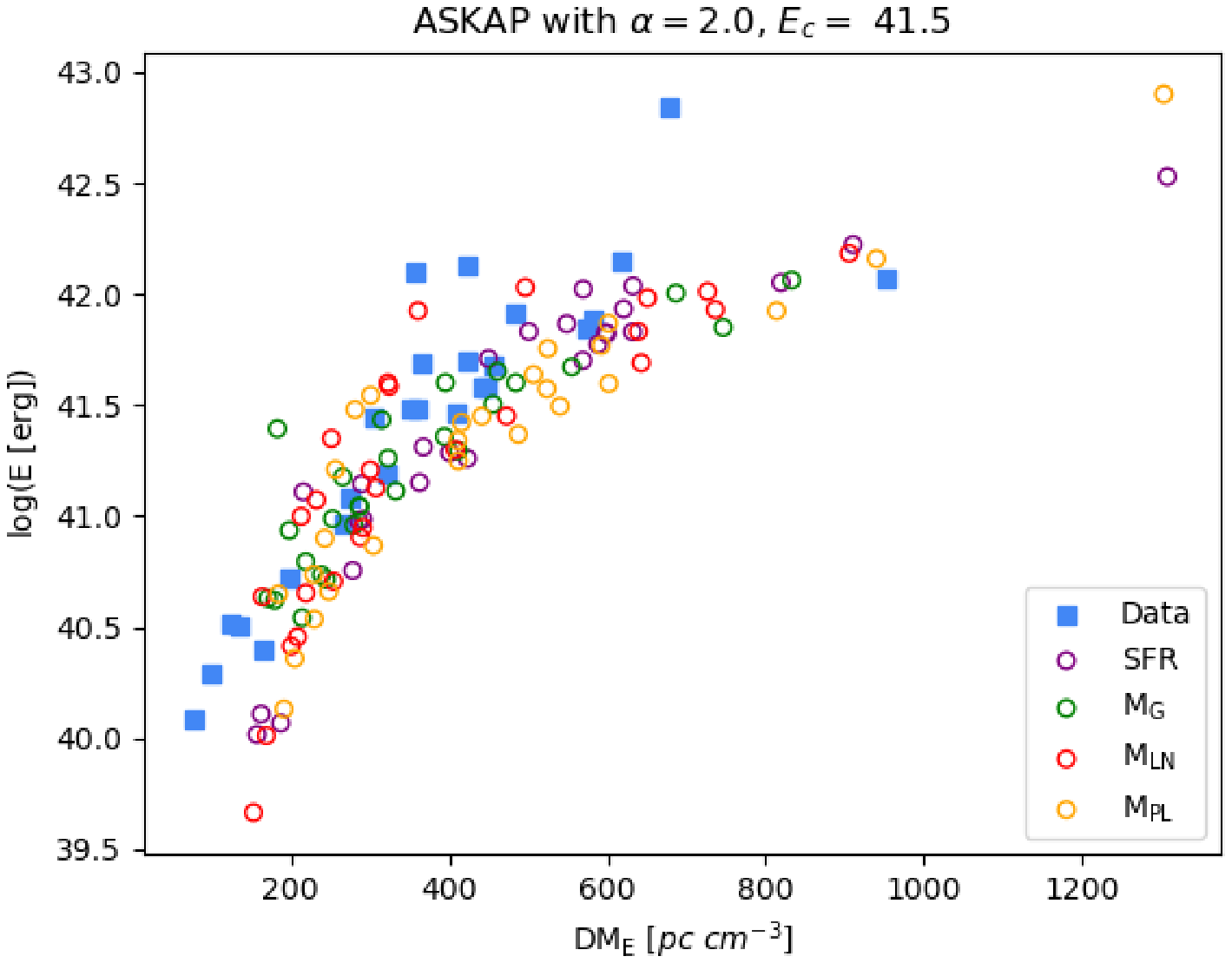}
    \includegraphics[width=\columnwidth]{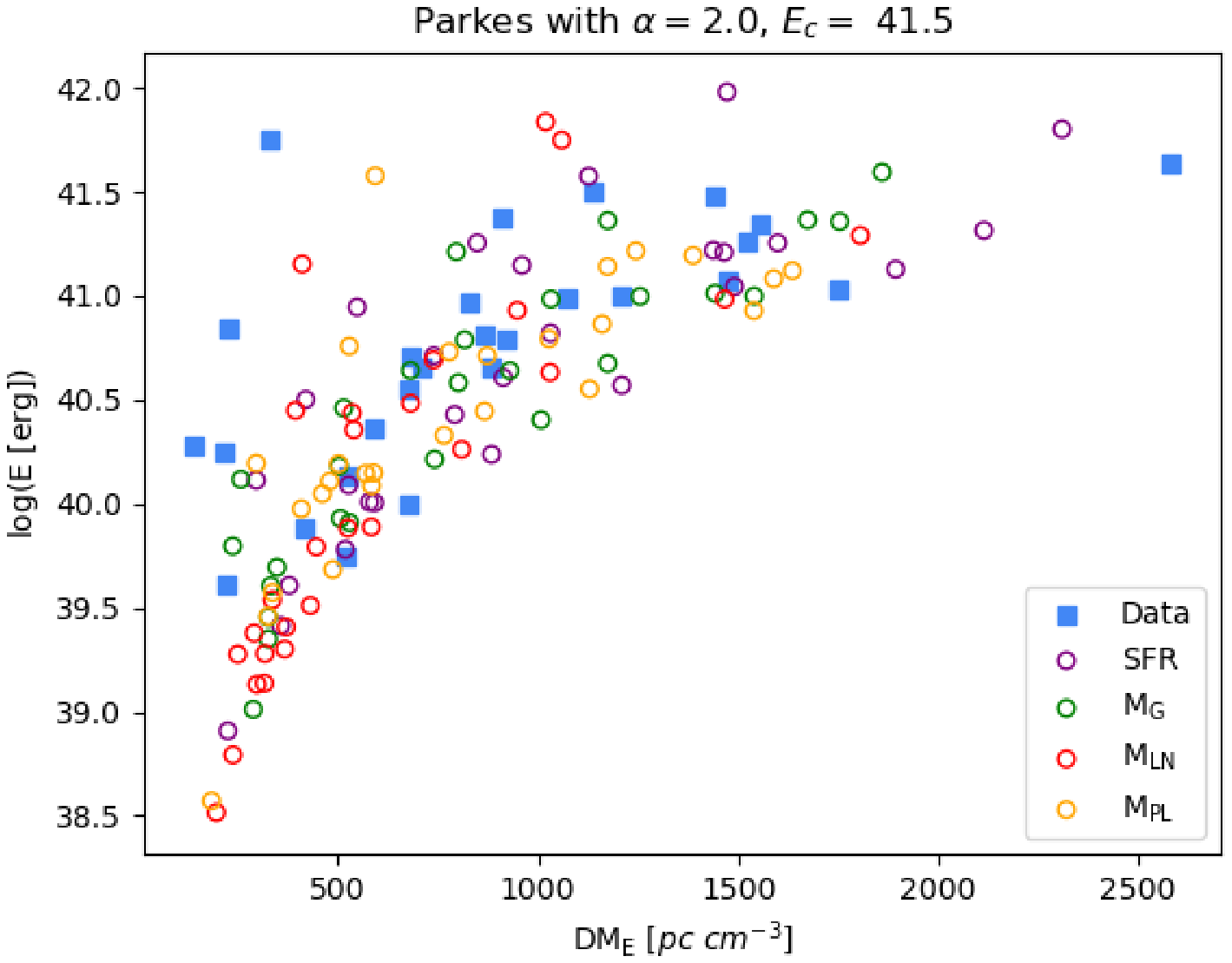}
    \caption{These six panels show the $\rm DM_E$ vs. Energy distribution plots of the data and four simulation models (notation is consistent with Fig.~\ref{fig:fig3}) for three $\alpha$ values of $1.6,$ $1.8,$ and $2.0$ and two subsamples of the data, the ASKAP and Parkes samples. A sample of each simulation was randomly selected to standardize the number of simulation and data events.}
    \label{fig:2d_dme_en}
\end{figure*}

    We apply the 2-D KS test to compare both $\rm DM_E$ and energy distributions between the data and the four simulation models. The last rows in Tables \ref{tab:diff alphas} and \ref{tab:diff ecut} shows the results of this test given their respective parameter setups. We also show the $\rm DM_E$ vs. Energy plots for the different powerlaw indices in Fig.~\ref{fig:2d_dme_en}. We can see that the 2-D tests do not reject additional models not rejected by 1-D tests.

To summarize our 1-D and 2-D tests, we  confirm that $\alpha=1.8$ describe the energy distribution of available FRB data very well. This is consistent with previous results obtained by independent groups \citep{luo18,luo20,lu19b,lu20b, lu20}. On the other hand, for $\alpha=1.8$, we cannot constrain $E_c$, in contrast to \cite{luo20}, suggesting that whether there exists an upper cutoff in the energy distribution is still an open question. Since all the redshift distribution models can pass the $\alpha=1.8$ model, we conclude that the current Parkes and ASKAP cannot offer definite clues regarding the intrinsic redshift distribution of FRBs.

\section{Conclusion and Discussion}\label{sec:conclusions}
In this paper, we perform a set of dedicated Monte Carlo simulations in an effort of constraining the energy and redshift distributions of FRBs. We confront two subsamples of FRBs detected by ASKAP and Parkes, respectively, with the simulation results assuming a cutoff power law energy distribution function and two redshift distribution models, one following the star formation rate history and the other tracking the compact binary mergers. For the latter, three merger delay timescale distributions models (Gaussian, lognormal, and power law) are considered. We reach the following conclusions: 
\begin{itemize}
\item We confirm that the isotropic-equivalent energy distribution of FRBs is well described by a power law function. The best-fit power law index is $\alpha=1.8$. The value $\alpha=2.0$ is acceptable, but $\alpha=1.6$ is disfavoured. This is consistent with the conclusions drawn by previous authors \cite{luo18,luo20,lu19b,lu20b,lu20}. 
\item While \cite{luo20} claimed a luminosity cutoff of $L_c \sim 3 \times 10^{44}$ erg/s, we were unable to constrain the cutoff energy $E_c$ in the energy distribution function with the current data. 
\item All of the redshift distribution models for $\alpha=1.8$ are not rejected for both the Parkes and ASKAP samples. As a result, the intrinsic redshift distribution of FRBs cannot be constrained with the current data. 
\end{itemize}

Our results can be compared with several previous papers. For the energy distribution, 
as we have shown in this paper (and also \citealt{luo18,luo20,lu19b,lu20b}), a power law energy distribution with index $\alpha=1.8$ gives a satisfactory description of both ASKAP and Parkes FRB samples. In terms of redshift distribution, \cite{cao18} claimed that the Parkes data may support a compact star merger redshift distribution model with a power law merger delay time distribution.  \cite{locatelli19} claimed that the ASKAP data may support a compact star merger redshift distribution, while Parkes may follow the star formation rate history redshift distribution. We were not able to confirm or deny either of their results with a larger Parkes sample and also an ASKAP sample. For $\alpha=1.8$, we find that the merger model with all three time delay distributions (Gaussian, log-normal, and powerlaw) is no more superior than the star formation rate model. 

Since we were unable to constrain the redshift distribution, the source/progenitor model(s) of repeating and non-repeating FRBs cannot be inferred. Better constraints on the intrinsic FRB redshift distribution models can be achieved using the methodology laid out in this paper, with a much larger, uniform FRB sample such as the upcoming CHIME sample. While the current FRB catalog has more than 100 FRBs, it is difficult to make use of every event that has been detected with different telescopes, due to the difficulties in reliably modeling vastly different observational selection effects. 

\section*{Acknowledgements}
We thank the referee for the helpful comments. RCZ acknowledges Emily Petroff for information on the FRB catalogue and Shunke Ai for sharing his redshift distribution simulation results for cross-comparisons. YL is supported by the KIAA-CAS Fellowship, which is jointly supported by Peking University and Chinese Academy of Sciences, and is partially supported by the China Postdoctoral Science Foundation (No. 2018M631242). DRL acknowledges support from the NSF awards OIA-1458952 and PHY-1430284.

\section*{Data Availability}
The simulation data underlying this article are available upon reasonable requests. 












\bsp	
\label{lastpage}
\end{document}